\documentclass[%
 reprint,
superscriptaddress,
 amsmath,amssymb,
 aps,
]{revtex4-2}

\usepackage{graphicx}% Include figure files
\usepackage{dcolumn}% Align table columns on decimal point
\usepackage{bm}% bold math
\begin{document}

\preprint{APS/123-QED}

\title{Propagation-invariant space-time supermodes in a multimode waveguide}% Force line breaks with \\

\author{Abbas Shiri}
\affiliation{CREOL, The College of Optics \& Photonics, University of Central Florida, Orlando, FL 32816, USA}
\affiliation{Department of Electrical and Computer Engineering, University of Central Florida, Orlando, FL 32816, USA}
\author{Scott Webster}
\affiliation{CREOL, The College of Optics \& Photonics, University of Central Florida, Orlando, FL 32816, USA}
\author{Kenneth L. Schepler}
\affiliation{CREOL, The College of Optics \& Photonics, University of Central Florida, Orlando, FL 32816, USA}
\author{Ayman F. Abouraddy}
\affiliation{CREOL, The College of Optics \& Photonics, University of Central Florida, Orlando, FL 32816, USA}
\affiliation{Department of Electrical and Computer Engineering, University of Central Florida, Orlando, FL 32816, USA}
\email{*raddy@creol.ucf.edu}

%\affiliation{%


%\affiliation{
%\affiliation{
%\affiliation{%


%\noindent\textbf{Keywords.} Space-time supermode, multi-mode waveguide optics, group-velocity, non-diffraction

\begin{abstract}
When an optical pulse is spatially localized in a highly multimoded waveguide, its energy is typically distributed among a multiplicity of modes, thus giving rise to a speckled transverse spatial profile that undergoes erratic changes with propagation. It has been suggested theoretically that pulsed multimode fields in which each wavelength is locked to an individual mode at a prescribed axial wave number will propagate invariantly along the waveguide at a tunable group velocity. In this conception, an initially localized field remains localized along the waveguide. Here, we provide proof-of-principle experimental confirmation for the existence of this new class of pulsed guided fields, which we denote `space-time supermodes', and verify their propagation invariance in a planar waveguide. By superposing up to 21 modes, each assigned to a prescribed wavelength, we construct space-time supermodes in a 170-micron-thick planar glass waveguide with group indices extending from $\approx1$ to $\approx2$. The initial transverse width of the field is 6~microns, and the waveguide length is 9.1~mm, which is $\approx257\times$ the associated Rayleigh range. A variety of axially invariant transverse spatial profiles are produced by judicious selection of the modes contributing to the ST supermode, including single-peak and multi-peak fields, dark fields (containing a spatial dip), and even flat uniform intensity profiles.
\end{abstract}

\maketitle

\section{Introduction}

Recent interest in multimode fibers has been driven by an anticipated data crunch in which the ever-increasing demand for high-bandwidth communication services may be thwarted by the limited capacity of single-mode fibers \cite{Mitra01N,Richardson10S}. Consequently, significant efforts have hitherto been directed to developing developing strategies for increasing data transmission rates via multimode fibers \cite{Ryf13OFC,Uden14NP,Sillard16JLT,Xiong18LSA, Pauwels19SR,Zhou21NC}. However, a multimode field in a fiber or waveguide does not maintain its transverse spatial profile. Indeed, when pulsed light is focused into a highly multimode waveguide, the field energy is typically distributed among a multiplicity of modes, resulting in the emergence of a random speckle pattern whose transverse spatial intensity profile undergoes erratic changes with propagation along the waveguide. %developing strategies for maintaining the field structure in a multimode waveguide; e.g., by exciting a single mode [] or stable supermodes \cite{Carpenter15NP}[].

A different approach to achieving \textit{propagation invariance} in a pulsed multimode waveguide field was suggested in \cite{Zamboni01PRE,Zamboni02PRE,Zamboni03PRE} that makes use of so-called X-waves. In free space, X-waves are superluminal propagation-invariant pulsed beams \cite{Lu92IEEEa,Saari97PRL,Reivelt03arxiv,Turunen10PO,FigueroaBook14}. By exciting a spatio-temporal field structure at the waveguide entrance that corresponds to a spectrally discretized X-wave, a pulsed supermode propagates invariantly along the waveguide \cite{Zamboni01PRE,Zamboni02PRE,Zamboni03PRE}. However, synthesizing X-waves requires ultrabroadband sources \cite{Saari96Conference,Grunwald03PRA}, the observed deviation in their group velocity from that of a conventional wave packet is extremely small in the paraxial regime \cite{Bonaretti09OE,Bowlan09OL,Kuntz09PRA}, and coupling them to a waveguide poses substantial experimental difficulties. As such, X-wave supermodes in multimode waveguides have to date remained theoretical entities. Additionally, the propagation of so-called focused-wave modes (FWMs \cite{Brittingham83JAP}) in single-mode fibers has been examined theoretically \cite{Vengsarkar92JOSAA,Ruano21JO}. However, the synthesis of FWMs in the optical domain suffers the same challenges as do X-waves \cite{Yessenov22AOP}.

The intrinsic challenges involved in synthesizing X-waves have been recently resolved by the devleopment of so-called `space-time' (ST) wave packets \cite{Kondakci16OE,Parker16OE,Yessenov19OPN,Yessenov22AOP}: propagation-invariant \cite{Kondakci17NP,Porras17OL,Efremidis17OL,Wong17ACSP2,Kondakci18OE,Wong20AS}, narrowband, paraxial pulsed beams whose group velocity can be tuned in free space \cite{Salo01JOA,Efremidis17OL,Wong17ACSP2,Kondakci19NC} or transparent dielectrics \cite{Bhaduri19Optica}. A host of novel phenomena have been recently demonstrated using ST wave packets including anomalous refraction \cite{Bhaduri20NP,AllendeMotz21OL}, ST Talbot effects \cite{Hall21APL} and dispersion-free propagation in dispersive media \cite{Orlov02OL,Porras03PRE2,Longhi04OL,Porras04PRE,Malaguti08OL,Malaguti09PRA,Yessenov21ACSPhot,Hall22LPR}, to name a few \cite{Yessenov22AOP}. The versatility of this new class of propagation-invariant pulsed beams suggests their suitability as a platform for synthesizing propagation-invariant supermodes in multimode waveguides. Indeed, discretizing the spectrum of a ST wave packet may yield propagation-invariant pulsed fields in a highly multimoded waveguides. We denote such field configurations `ST supermodes'. Recent theoretical studies have been directed along these lines, including investigations of linear and nonlinear generation of ST supermodes in planar multimode waveguides \cite{Guo21PRR} and in multimode fibers \cite{Kibler21PRL}, and the prediction of so-called helicon wave packets that incorporate orbital angular momentum into the ST supermode \cite{Bejot21ACSP}.

\begin{figure*}[t!]
\centering
\includegraphics[width=15.2cm]{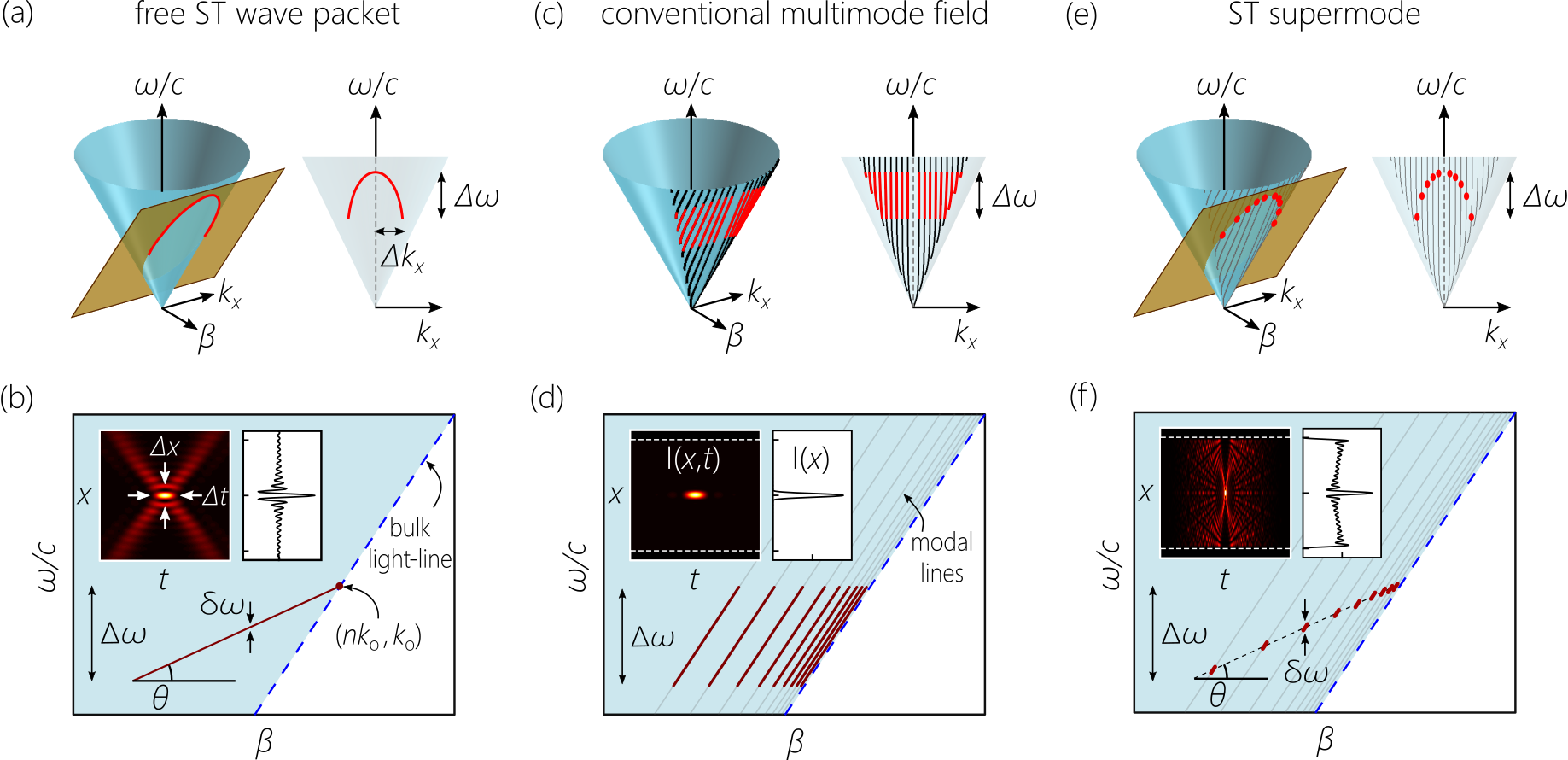}
\caption{\textbf{ST supermodes in a multimode waveguide.} (a) Representation in $(k_{x},\beta,\tfrac{\omega}{c})$-space of the spectral support domain on the surface of the light-cone of a freely propagating ST wave packet in a medium of refractive index $n$. The spectral support is the conic section (red curve) at the intersection of the light-cone with a tilted spectral plane. The spectral projection onto the $(k_{x},\tfrac{\omega}{c})$-plane is shown, and the spatial and temporal bandwidths $\Delta k_{x}$ and $\Delta\omega$, respectively, are identified. (b) The spectral projection for the ST wave packet from (a) onto the $(\beta,\tfrac{\omega}{c})$-plane is a straight line making an angle $\theta$ with the $\beta$-axis. The inset shows the X-shaped spatio-temporal intensity profile $I(x;t)$ and time-averaged intensity $I(x)\!=\!\int\!dtI(x;t)$ at $z\!=\!0$ [Fig.~\ref{Fig:Comparison}(a,b)]. (c) The spectral support for index-guided modes in a multimode planar waveguide of refractive index $n$ and thickness $d$ surrounded by free space. (d) The spectral projection onto the $(\beta,\tfrac{\omega}{c})$-plane for a multimode field of temporal bandwidth $\Delta\omega$; the energy is distributed among a multiplicity of waveguide modes. (e) The spectrally discretized support for an ST supermode associated with the same multimode waveguide from (c,d). (f) The spectral projection onto the $(\beta,\tfrac{\omega}{c})$-plane for a ST supermode is the set of points at the intersection of the spectral plane from (a,b) with the modal lines from (c,d). The full bandwidth is $\Delta\omega$, and each mode is assigned a particularly frequency and is associated with a narrow spectral uncertainty $\delta\omega\!\ll\!\Delta\omega$.}
\label{Fig:LightCones}
\end{figure*}

Here we formulate theoretically and verify experimentally the existence of this new class of propagation-invariant pulsed fields compatible with multimode waveguides that we call ST supermodes. An ideal ST supermode is a superposition of monochromatic waveguide modes that propagates invariantly at a prescribed group velocity, which can be tuned continuously across the subluminal and superluminal regimes. These surprising features of ST supermodes rely on careful selection of the wavelength and axial wave number of each mode in the superposition. In realistic ST supermodes, each mode is associated not with a single wavelength, but rather with a narrow spectral uncertainty that is nevertheless much smaller than the full supermode bandwidth. In our proof-of-principle experiments, we synthesize and launch ST supermodes comprising up to 21 modes in a planar waveguide in the form of a 170-$\mu$m-thick glass slide with free-space cladding and substrate at a wavelength of 800~nm. Using a spatio-temporal pulsed-beam shaper, we prepare in free space superpositions of waveguide modes assigned to prescribed optical frequencies, resulting in ST supermodes with tunable group velocity and characteristic X-shaped spatio-temporal profiles. Rather than the erratic speckle patterns resulting from a conventional superposition of the waveguide modes, the time-averaged intensity of the ST supermode is axially invariant; in other words, an initially localized ST supermode remains localized along the waveguide. The field is initially localized toa 6-$\mu$m-wide transverse intensity feature, so that the waveguide length corresponds to $\approx\!257\times$ the associated Rayleigh range. Moreover, by judicious selection of the modes contributing to the ST supermode, the time-averaged transverse intensity profile can be modified. We thus produce a variety of profiles, including ones with a single or multiple peaks, dark profiles (a central dip), and even flat uniform profiles.

\section{Theory of guided space-time supermodes}

\subsection{Free ST wave packets}

We start by briefly describing freely propagating ST wave packets, which can be visualized in terms of their spectral support domain on the surface of the light-cone. In a medium of refractive index $n$, the light-cone corresponds to the dispersion relationship $k_{x}^{2}+\beta^{2}\!=\!n^{2}(\tfrac{\omega}{c})^{2}$ [Fig.~\ref{Fig:LightCones}(a)], where $c$ is the speed of light in vacuum, $\omega$ is the temporal frequency, and $k_{x}$ and $\beta$ are the transverse and longitudinal components of the wave vector along the transverse and axial coordinates $x$ and $z$, respectively \cite{Donnelly93ProcRSLA,Yessenov19PRA,Yessenov22AOP}; we restrict ourselves in this proof-of-principle experiment to a single transverse coordinate for simplicity. This formulation is suitable for ST supermodes in a planar waveguide, and we discuss below the prospects for producing ST supermodes in conventional waveguides where the field is confined in both transverse coordinates.

The spectral support domain for a ST wave packet on the light-cone surface is a one-dimensional (1D) trajectory -- namely, the conic section at the intersection of the light-cone with a spectral plane $\beta\!=\!nk_{\mathrm{o}}+\Omega/\widetilde{v}$, where $\omega_{\mathrm{o}}$ is a fixed temporal frequency, $k_{\mathrm{o}}\!=\!\omega_{\mathrm{o}}/c$ the associated wave number, $\Omega\!=\!\omega-\omega_{\mathrm{o}}$, and $\widetilde{v}$ is its group velocity \cite{Kondakci17NP,Kondakci19NC}; see Fig.~\ref{Fig:LightCones}(a,b). This spectral plane is parallel to the $k_{x}$-axis, and makes an angle $\theta$ (the spectral tilt angle) with the $k_{z}$-axis, where $\widetilde{v}\!=\!c\tan{\theta}\!=\!c/\widetilde{n}$, and $\widetilde{n}\!=\!\cot{\theta}$ is the group index. Each frequency $\omega$ is thus associated with a single spatial frequency $\pm k_{x}$. The ST wave packet travels rigidly in the medium at a group velocity $\widetilde{v}$ that can in principle take on arbitrary values: subluminal $\widetilde{v}\!<\!c/n$ ($\omega_{\mathrm{o}}$ is the maximum frequency in the spectrum) or superluminal $\widetilde{v}\!>\!c/n$ ($\omega_{\mathrm{o}}$ is the minimum frequency) \cite{Bhaduri19Optica,Bhaduri20NP}. The ST wave packet has spatial and temporal bandwidths $\Delta k_{x}$ and $\Delta\omega$, respectively [Fig.~\ref{Fig:LightCones}(a)], and typically has a spatio-temporal profile $I(x;t)$ at every axial plane that is X-shaped with a transverse width $\Delta x\!\sim\!(\Delta k_{x})^{-1}$ at the pulse center $t\!=\!0$, and a pulse of width $\Delta T\!\sim\!(\Delta\omega)^{-1}$ at the beam center $x\!=\!0$ [Fig.~\ref{Fig:LightCones}(b), inset]. Other spatio-temporal profiles can be synthesized by modulating the spectral phase (e.g., Airy ST wave packets \cite{Kondakci18PRL} and ST caustics \cite{Wong21OE}). The time-averaged intensity $I(x)\!=\!\int\!dt I(x;t)$ takes the form of a localized spatial feature at $x\!=\!0$ atop a constant background [Fig.~\ref{Fig:LightCones}(b), inset].

The ideal scenario in which each frequency $\omega$ is associated with a single spatial frequency $k_{x}$ results in infinite-energy wave packets \cite{Sezginer85JAP}. In realistic scenarios, each $k_{x}$ is associated with a finite spectral uncertainty $\delta\omega\!\ll\!\Delta\omega$ [Fig.~\ref{Fig:LightCones}(b)]. In this case, the ST wave packet has finite energy, and also a finite propagation distance $L_{\mathrm{max}}\!\sim\!\tfrac{c}{\delta\omega|n(n-\cot{\theta})|}$ \cite{Yessenov19OE}. We have shown that the realistic ST wave packet can be decomposed into a product of the ideal ST wave packet traveling at $\widetilde{v}$ and a `pilot envelope' of temporal width $\sim\!(\delta\omega)^{-1}$ traveling at $c$. Axial walk-off between these two field structures occurs when the ST wave packet reaches the edge of the pilot envelope (when $z\!\sim\!L_{\mathrm{max}}$), whereupon the spatio-temporal X-shaped profile of the ST wave packet is deformed, and the field diffracts \cite{Kondakci16OE,Yessenov19OE}.

\begin{figure*}[t!]
\centering
\includegraphics[width=17.5cm]{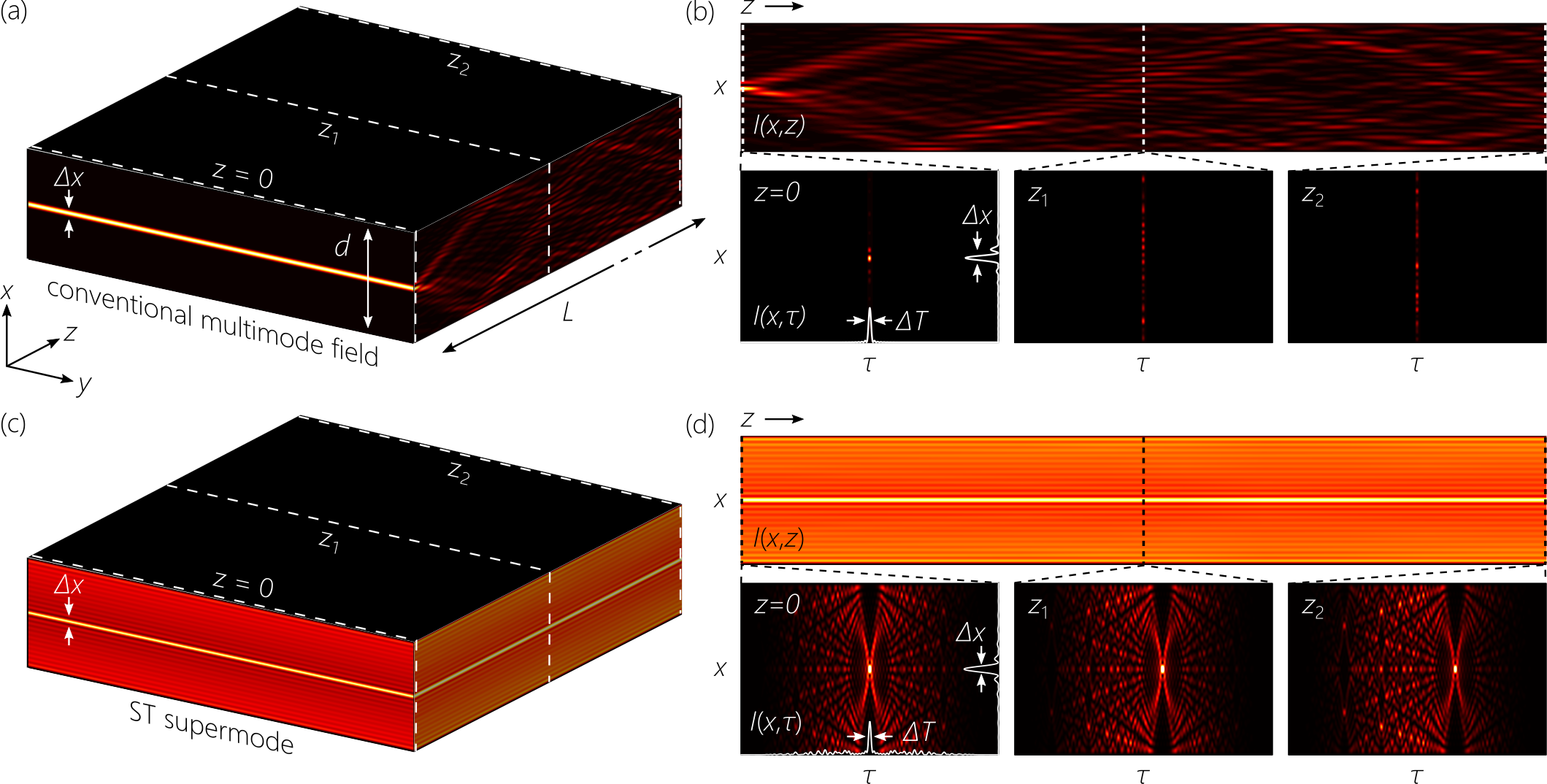}
\caption{\textbf{Conventional pulsed fields versus ST supermodes in a multimode waveguide.} (a,b) Calculated time-averaged intensity for a conventional pulsed multimode-waveguide field [Fig.~\ref{Fig:LightCones}(c,d)] localized at the waveguide entrance. The initially localized field spreads rapidly across the waveguide width, and a speckle pattern is formed that undergoes erratic changes with propagation. We also plot the spatio-temporal intensity profile $I(x;\tau)$ at $z\!=\!0$ and at two axial positions $z_{1}$ and $z_{2}$. Here, $\tau\!=\!t-nz/c$ is in the traveling frame of the wave packet. (c,d) Same as (a,b), but for a ST supermode [Fig.~\ref{Fig:LightCones}(e,f)]. The time-averaged intensity profile is maintained along the waveguide, and the spatio-temporal profile at $z\!=\!0$. The ST supermode is designed to travel at a subluminal group velocity $\widetilde{v}$ (slower than the conventional pulsed multimode field). The dimensions of the waveguide and the parameters of the initial field structure correspond to those in our experiments, but the main features of these simulations are generic. The calculations in (a,b) correspond to Fig.~\ref{Fig:NumberOfModes}(a), and those in (c,d) correspond to Fig.~\ref{Fig:NumberOfModes}(d).}
\label{Fig:Comparison}
\end{figure*}

\subsection{Conventional multimode-waveguide field}

Consider the field in a multimode planar waveguide of refractive index $n$ and thickness $d$, with free space on both sides. The spectral representation for each guided mode is a 1D curve on the surface of the light-cone [Fig.~\ref{Fig:LightCones}(c)]. When a pulse of bandwidth $\Delta\omega$ and finite spatial width $\Delta x\!<\!d$ is launched into the waveguide, its energy is typically distributed among a multiplicity of modes. If the input field at $z\!=\!0$ is $E(x,z\!=\!0;t)\!=\!e^{-i\omega_{\mathrm{o}}t}\psi_{x}(x)\psi_{t}(t)$, where $\omega_{\mathrm{o}}$ is the carrier frequency, $\psi_{x}(x)$ is the initial transverse spatial profile, and $\psi_{t}(t)$ is the initial temporal profile, then the multimode-waveguide field can be written as:
\begin{equation}\label{Eq:ConventionalWGField}
E_{\mathrm{WG}}(x,z;t)=e^{-i\omega_{\mathrm{o}}t}\sum_{m}A_{m}e^{i\beta_{m}z}u_{m}(x)\psi_{m}(z;t);
\end{equation}
here, $u_{m}(x)$ is the spatial profile of the $m^{\mathrm{th}}$-mode (integer $m$ is the modal index), which we take to be approximately wavelength-independent (in light of the small bandwidth employed in our experiments), $\beta_{m}$ its axial wave number estimated at $\omega_{\mathrm{o}}$, $A_{m}\!=\!\int\!dxu_{m}^{*}(x)\psi_{x}(x)$, and $\psi_{m}(z;t)$ is an axial envelope:
\begin{equation}\label{Eq:AxialModalEnvelope}
\psi_{m}(z,t)=\int\!d\Omega\widetilde{\psi}_{t}(\Omega)e^{-i\Omega(t-z/\widetilde{v}_{m})}e^{i\beta_{2m}\Omega^{2}z/2},
\end{equation}
where $\widetilde{v}_{m}$ and $\beta_{2m}$ are the group velocity and GVD coefficient for the $m^{\mathrm{th}}$-mode evaluated at $\omega_{\mathrm{o}}$, respectively. The time-averaged intensity $I_{\mathrm{WG}}(x,z)\!=\!\int\!dt|E_{\mathrm{WG}}(x,z;t)|^{2}$ is:
\begin{eqnarray}
I_{\mathrm{WG}}(x,z)=\sum_{m}|A_{m}|^{2}|u_{m}(x)|^{2} &+&\nonumber\\
\!\!\!\!\!2\!\!\!\!\!\!\sum_{m\neq n,m>n}\!\!\!\!\!|A_{m}A_{n}u_{m}(x)u_{n}(x)f_{mn}(z)|\cos{\{\chi_{mn}(x,z)\}},
\end{eqnarray}
where the function $f_{mn}(z)\!=\!|f_{mn}(z)|e^{i\xi_{mn}(z)}$ is defined as:
\begin{equation}
f_{mn}(z)\!=\!\int\!d\Omega|\widetilde{\psi}_{t}(\Omega)|^{2}e^{-i(\widetilde{v}_{m}^{-1}-\widetilde{v}_{n}^{-1})\Omega z}e^{-i(\beta_{2m}-\beta_{2n})\Omega^{2}z/2}.
\end{equation}
By defining the complex field quantities as $A_{m}\!=\!|A_{m}|e^{i\alpha_{m}}$ and $u_{m}(x)\!=\!|u_{m}(x)|e^{i\phi_{m}(x)}$, we have $\chi_{mn}(x,z)\!=\!\varphi_{m}(x,z)-\varphi_{n}(x,z)-\xi_{mn}(z)$, where $\varphi_{m}(x,z)\!=\!\phi_{m}(x)+\beta_{m}z+\alpha_{m}$.

Consequently, the spectral support domain for this conventional pulsed multimode field in a planar waveguide is a collection of 1D trajectories, all potentially of bandwidth $\Delta\omega$ [Fig.~\ref{Fig:LightCones}(c,d)]. The field is a superposition of waveguide modes, but the field collectively is \textit{not} propagation invariant. Indeed, an initially localized field at the waveguide entrance $z\!=\!0$ spreads spatially along the transverse dimension [Fig.~\ref{Fig:Comparison}(a)], and typically displays a speckle structure whose intensity distribution undergoes irregular changes with propagation [Fig.~\ref{Fig:Comparison}(b)].

\subsection{Ideal ST supermodes}

ST supermodes present an altogether different picture from that of conventional pulsed multimode fields. An ST supermode is a pulsed field in a multimode waveguide whose spectral support domain is the intersection of those for two distinct field configurations: (1) the spectral support associated with a conventional multimode waveguide field, which is the collection of 1D trajectories on the light-cone surface corresponding to the waveguide modes [Fig.~\ref{Fig:LightCones}(c,d)]; and (2) the spectral support for an ST wave packet, which is the conic section at the intersection of the light-cone with a spectral plane [Fig.~\ref{Fig:LightCones}(a,b)]. The resulting spectral support domain for an ST supermode is therefore a set of points [Fig.~\ref{Fig:LightCones}(e,f)], each representing a monochromatic mode: the $m^{\mathrm{th}}$-mode is associated with a frequency $\omega_{\mathrm{m}}$ and an axial wave number $\beta_{m}(\omega_{m})\!=\!nk_{\mathrm{o}}+\tfrac{\omega_{m}-\omega_{\mathrm{o}}}{\widetilde{v}}$. The total temporal bandwidth is $\Delta\omega$, but the spectrum is discretized at frequencies $\omega_{m}$ [Fig.~\ref{Fig:LightCones}(f)]. The ST supermode field can be written as $E_{\mathrm{ST}}(x,z;t)\!=\!e^{i(\beta_{\mathrm{o}}z-\omega_{\mathrm{o}}t)}\psi_{\mathrm{ST}}(x,z;t)$, where the envelope is:
\begin{eqnarray}\label{Eq:STSupermodeIdeal}
\psi_{\mathrm{ST}}(x,z;t)&=&\sum_{m}A_{m}u_{m}(x)e^{-i(\omega_{m}-\omega_{\mathrm{o}})(t-z/\widetilde{v})}\nonumber\\&=&
\psi_{\mathrm{ST}}(x,0;t-z/\widetilde{v});
\end{eqnarray}
that is, the envelope travels rigidly at a group velocity $\widetilde{v}$. The spatio-temporal profile of the ST supermode [Fig.~\ref{Fig:LightCones}(f), inset] has a central X-shaped feature similar to its free-space counterpart [Fig.~\ref{Fig:LightCones}(b), inset], which travels at the group velocity $\widetilde{v}$. This group velocity is essentially independent of the group velocities $\widetilde{v}_{m}$ of the underlying modes. Because of spectral discretization \cite{Hall21APL}, the axial profile of the ST supermode extends beyond the central X-shaped feature associated with a freely propagating continuous-spectrum ST wave packet [Fig.~\ref{Fig:LightCones}(a,b)]. We discuss below the temporal extent of the ST supermode once we introduce the associated spectral uncertainty.

Unlike a conventional pulsed multimode field [Fig.~\ref{Fig:LightCones}(c,d) and Fig.~\ref{Fig:Comparison}(a,b)] that evolves erratically along the propagation axis, the time-averaged intensity of the ST supermode is given by:
\begin{equation}\label{Eq:TimeAveragedIntensity}
I_{\mathrm{ST}}(x,z)=\int\!dt|E_{\mathrm{ST}}(x,z;t)|^{2}=\sum_{m}|A_{m}|^{2}|u_{m}(x)|^{2},
\end{equation}
which is altogether independent of $z$ [Fig.~\ref{Fig:Comparison}(c,d)]. This is a general statement that is independent of the details of the waveguide and of the superposition weights. In general, the time-averaged intensity takes the form of a central localized spatial feature atop a constant background (which, however, can be purposefully modified as shown in Fig.~\ref{Fig:ModalShape} below). 

\subsection{Realistic ST supermodes in presence of spectral uncertainty}

In practice, a realistic ST supermode will be a superposition of modes centered at the frequencies $\omega_{m}$, but have finite spectral uncertainties $\delta\omega\!\ll\!\Delta\omega$ rather than being ideally monochromatic [Fig.~\ref{Fig:LightCones}(e,f)], so that the field envelope becomes:
\begin{equation}\label{Eq:STSupermodeWithUncertainty}
\psi_{\mathrm{ST}}(x,z;t)=\sum_{m}A_{m}u_{m}(x)e^{-i(\omega_{m}-\omega_{\mathrm{o}})(t-z/\widetilde{v})}\psi_{m}(z;t).
\end{equation}
The field combines features from the conventional pulsed multimode field (Eq.~\ref{Eq:ConventionalWGField}) and that of the ideal ST supermode (Eq.~\ref{Eq:STSupermodeIdeal}). The axial envelope $\psi_{m}(z;t)$ in Eq.~\ref{Eq:STSupermodeWithUncertainty} has the same form as in Eq.~\ref{Eq:AxialModalEnvelope}, but with one crucial difference: the integration is carried out for each mode over a small bandwidth $\delta\omega$ centered at $\omega_{m}$ (rather than over the full bandwidth $\Delta\omega$), so that the impact of GVD is significantly reduced. Assuming a Gaussian spectral profile for each mode $\widetilde{\psi}_{m}(\Omega)\!\propto\!\exp\{-\tfrac{\Omega^{2}}{2(\delta\omega)^{2}}\}$ to replace $\widetilde{\psi}_{t}(\Omega)$, we have:
\begin{equation}
\psi_{m}(z;t)\!\propto\!\exp\left\{-\frac{(t-z/\widetilde{v}_{m})^{2}}{1+(z/L_{m})^{2}}\frac{(\delta\omega)^{2}}{2}\left(1+i\mathrm{sgn}(\beta_{2m})\frac{z}{L_{m}}\right)\right\};
\end{equation}
here $\Omega\!=\!\omega-\omega_{m}$, $\mathrm{sgn}(\beta_{2m})$ is the sign of the GVD coefficient, and $L_{m}\!=\!|\beta_{2m}|(\delta\omega)^{2}$ is the dispersion length. Each axial envelope $\psi_{m}(z;t)$ can be viewed as a `pilot envelope' traveling at a group velocity $\widetilde{v}_{m}$ and having a dispersion length $L_{m}$. The central X-shaped spatio-temporal feature of the ST supermode travels at the tunable group velocity $\widetilde{v}$, which is essentially independent of $\widetilde{v}_{m}$, but the \textit{axial} extent of the spatio-temporal profile is determined by $\psi_{m}(z;t)$. Axial walk-off thus occurs between the X-shaped feature and the overall pilot envelope, reaching the edge of the envelope when $z\!\sim\!\tfrac{c}{\delta\omega|n(n-\widetilde{n})|}$.

However, the propagation of a ST supermode offers a surprising contrast to that of a freely propagating ST wave packet. The time-averaged intensity of the realistic ST supermode once again takes the form $I_{\mathrm{ST}}(x,z)\!=\!\sum_{m}|A_{m}|^{2}|u_{m}(x)|^{2}$, which is independent of $z$, just as in the case of the ideal ST supermode (Eq.~\ref{Eq:TimeAveragedIntensity}), as long as the spectral support domains of the underlying modes do \textit{not} overlap spectrally. In other words, even after the central X-shaped feature has disappeared beyond the edge of the pilot envelopes $\psi_{m}(z;t)$, the time-averaged intensity remains stable. In contrast, a freely propagating ST wave packet diffracts after this axial distance \cite{Yessenov19OE,Bhaduri19Optica}.

\section{Experiment}

\subsection{Waveguide sample}

\begin{figure*}[t!]
\centering
\includegraphics[width=17.5cm]{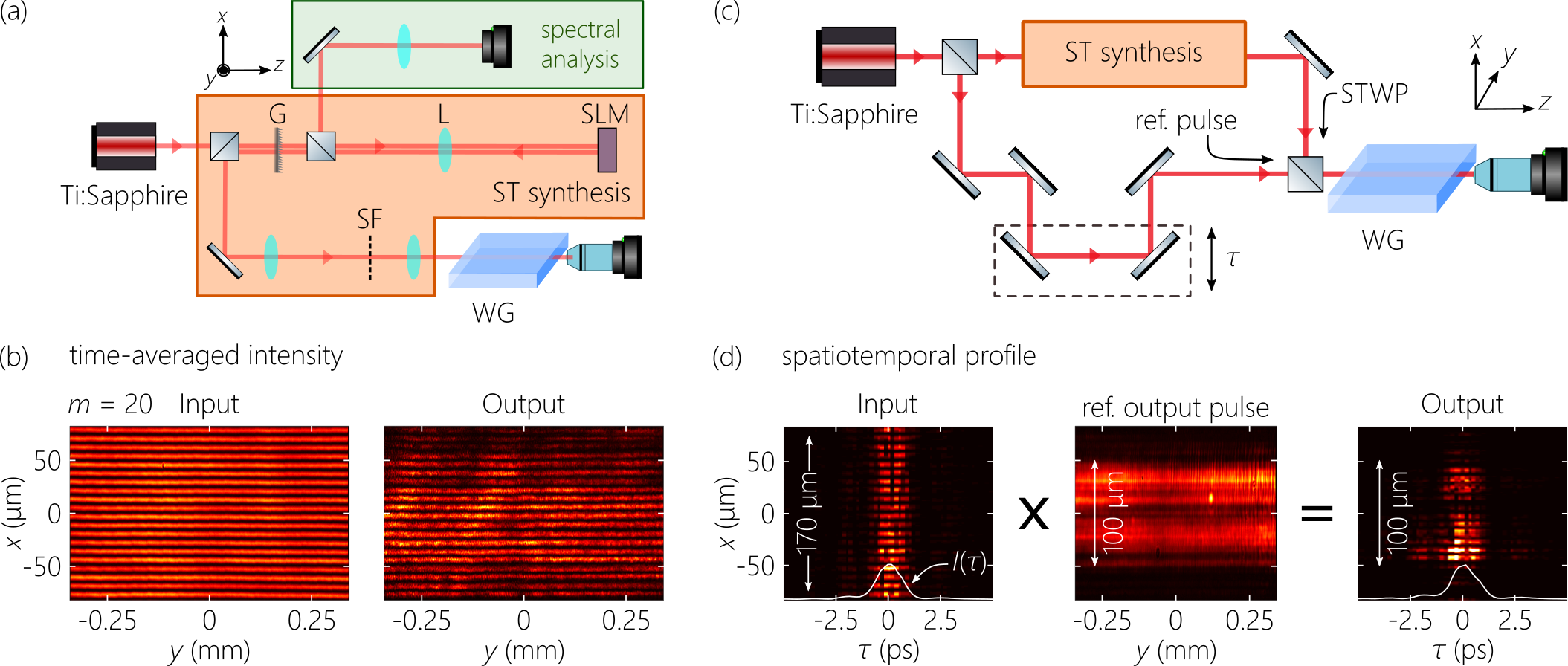}
\caption{\textbf{Experimental arrangement for excitation and characterization of individual waveguide modes and ST supermodes.} (a) Schematic of the arrangement for synthesizing free-space ST wave packets, acquiring the spatio-temporal spectrum, and coupling to the waveguide. G: Diffraction grating; L: cylindrical lens; SLM: spatial light modulator; SF: spatial filter. (b) Measured time-averaged intensity at the entrance and exit of the waveguide for the $m\!=\!20$ TM mode. (c) Setup for re-constructing the spatio-temporal intensity profile of a wave packet. (d) Reconstructed spatio-temporal intensity profile for the pulsed $m\!=\!20$ TM mode at the entrance (left panel) and exit (right panel) of the waveguide, in addition to the time-averaged intensity at the waveguide exit for the reference pulse (middle panel). The white curves at the bottom of the left and right panels are $I(\tau)\!=\!\int\!dxI(x;\tau)$, and $\tau\!=\!t-nz/c$ is measured in the frame of the propagating conventional wave packet.}
\label{Fig:Setup}
\end{figure*}

We make use of a planar waveguide in the form of a borosilicate glass slide (Schott D263) of refractive index $n\!=\!1.51$ at $\lambda\!=\!800$~nm, which is held in free space so that the substrate and cladding have $n\!=\!1$. The dimensions of the waveguide are as follows: thickness $d\!\approx\!170$~$\mu$m along $x$; length $L\!\approx\!9.1$~mm along $z$; and width $W\!\approx\!2$~mm along $y$. Optical-quality facets are produced by cleaving. The waveguide is highly multimoded with $M\!=\!\tfrac{2d}{\lambda}\!\sim\!470$ TM modes at $\lambda\!\sim\!800$~nm. The transverse spatial profiles of the low-order modes are practically identical to the corresponding modes in a waveguide formed of two perfect mirrors sandwiching a layer of the same thickness $d$ and refractive index $n$, which are sinusoids having a single transverse wave number $m\tfrac{\pi}{d}$. Significant deviation between these two classes of modes occurs only for higher-order modes ($>\!99\%$ overlap for $m\!<\!45$).

\subsection{ST wave-packet synthesis in free space}

The experimental arrangement for synthesizing ST supermodes and coupling them to the waveguide is depicted in Fig.~\ref{Fig:Setup}(a). We first spatially expand pulses from a Ti:sapphire laser (Coherent, MIRA; pulsewidth $\Delta T\!\approx\!150$~fs, bandwidth $\Delta\lambda\!\approx\!6$~nm) to extend over 25~mm. The pulse spectrum is resolved spatially via a diffraction grating (1200~lines/mm), the wavefront corresponding to the second diffraction order is collimated via a cylindrical lens (focal length $f\!=\!500$~mm), and is directed to a reflective, phase-only spatial light modulator (SLM; Hamamatsu X10468-02). The SLM intercepts a portion of the spectrally resolved wavefront corresponding to a bandwidth $\Delta\lambda\!\approx\!1.7$~nm (pulsewidth $\Delta T\!\approx\!1.5$~ps).

A portion of the wavefront retro-reflected from the SLM undergoes a spatial Fourier Transform implemented via a lens, and then directed to a CCD camera [Fig.~\ref{Fig:Setup}(a)]. This reveals the spatio-temporal spectrum projected onto the $(k_{x},\lambda)$-plane and verifies that the targeted spectral structure is synthesized. This spectrum takes the approximate form of a parabola $\tfrac{\Omega}{\omega_{\mathrm{o}}}\!=\!\tfrac{k_{x}^{2}}{k_{\mathrm{o}}^{2}}\tfrac{1}{2|1-\widetilde{n}_{\mathrm{a}}|}$ \cite{Kondakci19NC,Bhaduri20NP}, where $\widetilde{n}_{\mathrm{a}}\!=\!\cot{\theta_{\mathrm{a}}}$ is the group index of the ST field in free space and $\theta_{\mathrm{a}}$ is the free-space spectral tilt angle. We can estimate $\theta_{\mathrm{a}}$ from the curvature of the measured spatio-temporal spectrum. The group index $\widetilde{n}\!=\!\cot{\theta}$ of the ST supermode in the waveguide of refractive index $n$ is related to the free-space value via the following relationship \cite{Bhaduri20NP}:
\begin{equation}\label{Eq:LawOfRefraction}
1-\widetilde{n}_{\mathrm{a}}=n(n-\widetilde{n}).
\end{equation}
We have ignored here chromatic dispersion in the waveguide material, which can be easily accounted by by a modification of Eq.~\ref{Eq:LawOfRefraction}; see \cite{He21arxiv,Yessenov22OL}.

The wavefront retro-reflected from the SLM returns to the grating where the ST wave packet is formed. A $4f$ imaging system incorporating a spatial-frequency stop-filter in the Fourier plane eliminates the unscattered field component. The synthesized optical field delivered to the waveguide entrance has a width $\approx\!170$~$\mu$m along $x$, thus not requiring a coupling lens. The polarization of the field corresponds to a TM waveguide mode.

\subsection{Launching individual waveguide modes}

We confirm in Fig.~\ref{Fig:Setup}(b) the ability to excite individual waveguide modes using our field-synthesis arrangement. Each wavelength $\lambda$ in the spectrally resolved wavefront occupies a column in the SLM active area, along which we impart a phase distribution corresponding to a prescribed spatial frequency $k_{x}(\lambda)$. To excite a single waveguide mode, we hold $k_{x}\!=\!\pm m\tfrac{\pi}{d}$ fixed for all $\lambda$; i.e., the SLM phase pattern is independent of $\lambda$. We plot in Fig.~\ref{Fig:Setup}(b) the measured transverse intensity profiles $I_{m}(x,y)$ for the TM mode $m\!=\!20$ at the waveguide entrance and exit, verifying the expected propagation invariance of individual modes. The time-averaged intensity of the ST wave packet at the entrance or exit of the waveguide is captured by a CCD camera (The ImagingSource, DMK 27BUP031) equipped with an objective lens (Olympus PLN $20\times$).

\subsection{Time-resolved measurements}

We reconstruct the spatio-temporal profile of the ST supermodes by modifying the interferometric technique used in our previous work for freely propagating ST wave packets \cite{Yessenov19OE,Bhaduri20NP}. We co-launch the ST wave packet ($\Delta\lambda\!\approx\!1.7$~nm and $\Delta T\!\approx\!1.5$~ps) into the waveguide along with a plane-wave reference pulse from the initial laser ($\Delta\lambda\!\approx\!6$~nm and $\Delta T\!\approx\!150$~fs); see Fig.~\ref{Fig:Setup}(c). The ST supermode travels along the waveguide at a tunable group velocity $\widetilde{v}\!=\!c\tan{\theta}$, whereas the reference pulse travels at $\widetilde{v}_{\mathrm{ref}}\!=\!c/\widetilde{n}_{\mathrm{ref}}$, where $\widetilde{n}_{\mathrm{ref}}$ is the group index of the reference pulse in the waveguide. Because $d\!\gg\!\lambda$ and the reference pulse is coupled predominantly to the fundamental mode, $\widetilde{n}_{\mathrm{ref}}\!\approx\!n\!\approx\!1.51$. When the two pulsed fields overlap in space and time at the CCD at the waveguide exit, spatially resolved interference fringes are observed. By scanning an optical delay $\tau$ in the path of the reference pulse [Fig.~\ref{Fig:Setup}(c)], the intensity profile of the ST wave packet can be reconstructed from the visibility of the interference fringes. By repeating the measurements at the waveguide input, we can estimate the relative group delay between the ST supermode and the reference pulse incurred upon traversing the waveguide. Finally, using this arrangement, we can also estimate the group velocity of the ST supermode in free space $\widetilde{v}_{\mathrm{a}}\!=\!c\tan{\theta_{\mathrm{a}}}$. This measurement can then be compared to the value of the free-space spectral tilt angle $\theta_{\mathrm{a}}$ estimated from the measured spatio-temporal spectrum [Fig.~\ref{Fig:Setup}(a)].

The measured spatio-temporal profile of the $m\!=\!20$ waveguide mode is reconstructed in Fig.~\ref{Fig:Setup}(d), which reveals that the modal profile is separable with respect to the spatial and temporal degrees of freedom. The measurements in Fig.~\ref{Fig:Setup}(d) highlight a drawback of our approach for reconstructing the spatio-temporal profiles of the ST supermodes. Because the reference pulse is coupled into the fundamental waveguide mode (to avoid transverse spatial field variations), its intensity drops in the vicinity of the waveguide walls. We hence cannot reconstruct the profile of the ST supermode in those areas, and we are confined to a width of $\approx\!100$~$\mu$m of the full 170-$\mu$m-wide waveguide.

\section{Excitation and characterization of Space-time supermodes}

\subsection{Conventional multimode field}

\begin{figure*}[t!]
\centering
\includegraphics[width=16.5cm]{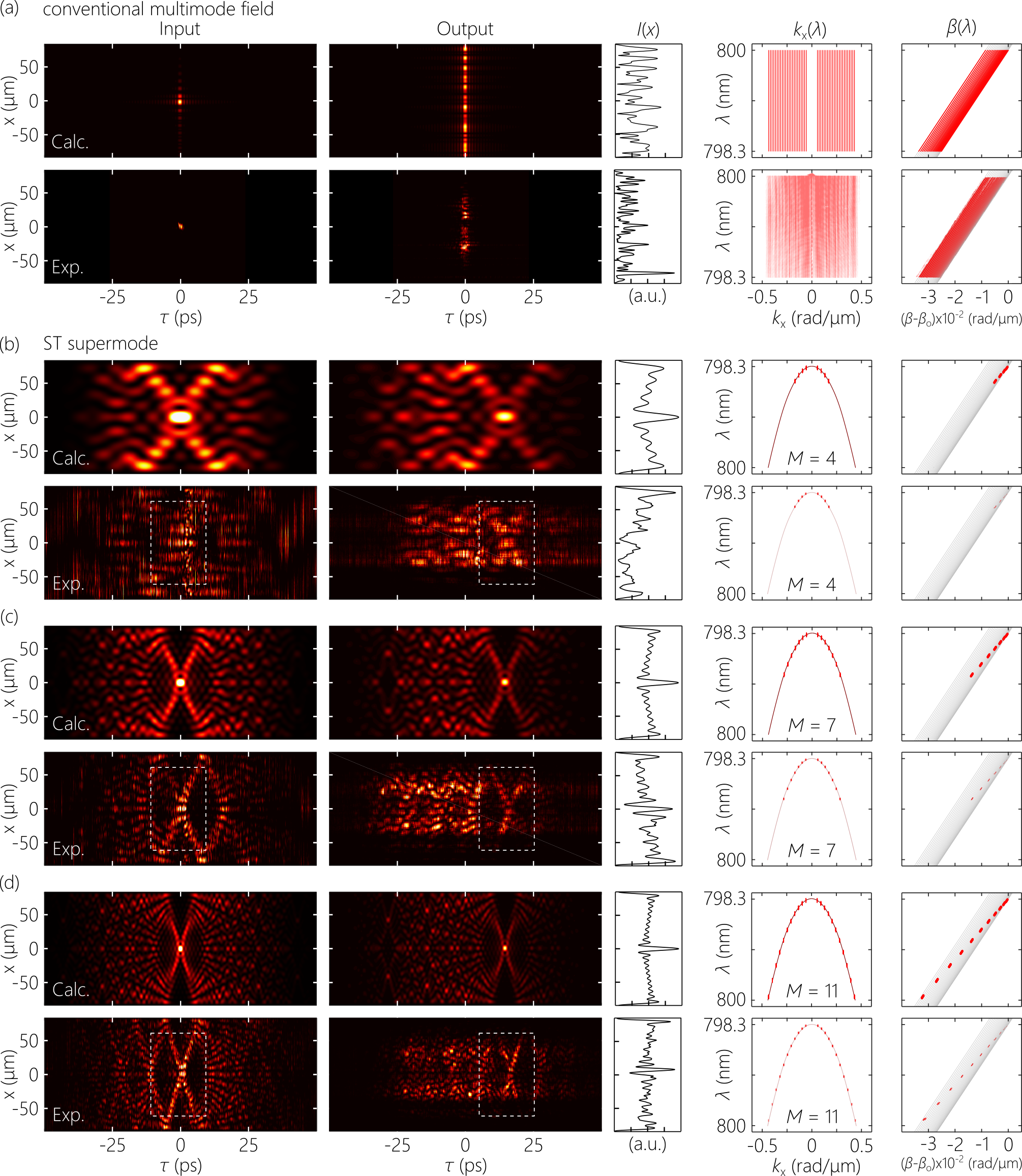}
\caption{\textbf{Observation of ST supermodes in a multimode waveguide.} Columns from left to right: input spatio-temporal intensity profile $I(x,z=0;\tau)$, where $\tau\!=\!t-nz/c$; output spatio-temporal intensity profile $I(x,z=L;\tau)$; time-averaged output intensity $I(x)$; spatio-temporal spectrum $k_{x}(\lambda)$; and dispersion relationship $\beta(\lambda)$. (a) A $\Delta T\!\approx\!1.5$-ps pulse is focused along $x$ to form a beam of transverse spatial width $\Delta x\!\approx\!6$~$\mu$m at the waveguide input couples to a multiplicity of modes, and the field as a result spreads across the width of the waveguide before reaching the exit. The first row is the calculated (`Calc.') result and the second row is the experimental (`Exp.') result. (b-d) Gradual formation of a subluminal ST supermode with $\theta\!=\!26.6^{\circ}$ ($\theta_{\mathrm{a}}\!=\!30^{\circ}$). Increasing the number of even-parity modes selected results in the gradual emergence of the characteristic X-shaped central feature of the ST supermode. (b) The ST supermode comprises $M\!=\!4$~modes, (c) 7~modes, and (d) 11~modes, the latter resulting in $\Delta x\!\approx\!6$~$\mu$m.}
\label{Fig:NumberOfModes}
\end{figure*}

As a basis for comparison, we examine the propagation of a conventional pulsed multimode field along the waveguide. On the SLM we place a quadratic phase pattern that produces a $\approx\!60$-$\mu$m-wide focused field along $x$. The $4f$ imaging system after the diffraction grating further reduces the beam width by $10\times$, resulting in a beamwidth $\Delta x\!\approx\!6$~$\mu$m at the waveguide entrance. The calculated and measured spatio-temporal profiles at the waveguide input and output are plotted in Fig.~\ref{Fig:NumberOfModes}(a). The initial spatio-temporal profile is separable in space and time, with temporal linewidth $\Delta T\!\approx\!1.5$~ps. The initially localized profile along $x$ spreads across the waveguide width before reaching the exit. The time-averaged intensity $I(x)$ takes the form of a speckle pattern. Although a multiplicity of modes are excited, the waveguide length and modal dispersion are not sufficiently large to produce temporal spreading. The pulse width at the waveguide output is thus approximately equal to that at the entrance. The acquired spatio-temporal spectrum $k_{x}(\lambda)$ is separable with respect to $k_{x}$ and $\lambda$, with spatial and temporal bandwidths $\Delta k_{x}\!\approx\!0.48$~rad/$\mu$m, and $\Delta\lambda\!\approx\!1.7$~nm, respectively.  

\subsection{Spectral discretization and formation of ST supermodes}

To excite an ST supermode, we superpose waveguide modes, each associated with a single wavelength, by modifying the 2D phase distribution imparted by the SLM to the impinging spectrally resolved wavefront. Because each wavelength $\lambda$ is incident along a different SLM column, we impart the transverse wave number $k_{x}(\lambda)\!=\!\pm m\tfrac{\pi}{d}$ associated with the approximate waveguide mode $u_{m}(x)$ to the selected wavelength. The bandwidth intercepted by the SLM is $\Delta\lambda\!\approx\!1.7$~nm spread across 800~pixels ($\approx\!2.1$~pm/pixel). However, the spectral resolution of the grating is $\approx\!14$~pm, corresponding to $\approx\!5$~pixels. We associate $21$~SLM pixels (spectral uncertainty $\delta\lambda\!\approx\!42$~pm, corresponding to a pulsewidth $\Delta T\!\approx\!44$~ps) with each mode. Therefore, in the process of synthesizing ST supermodes, the field spectrum is discretized. We assign zero-phase at the SLM to the unwanted wavelengths, which are thereby eliminated via the Fourier spatial filter.

\begin{figure*}[t!]
\centering
\includegraphics[width=16.8cm]{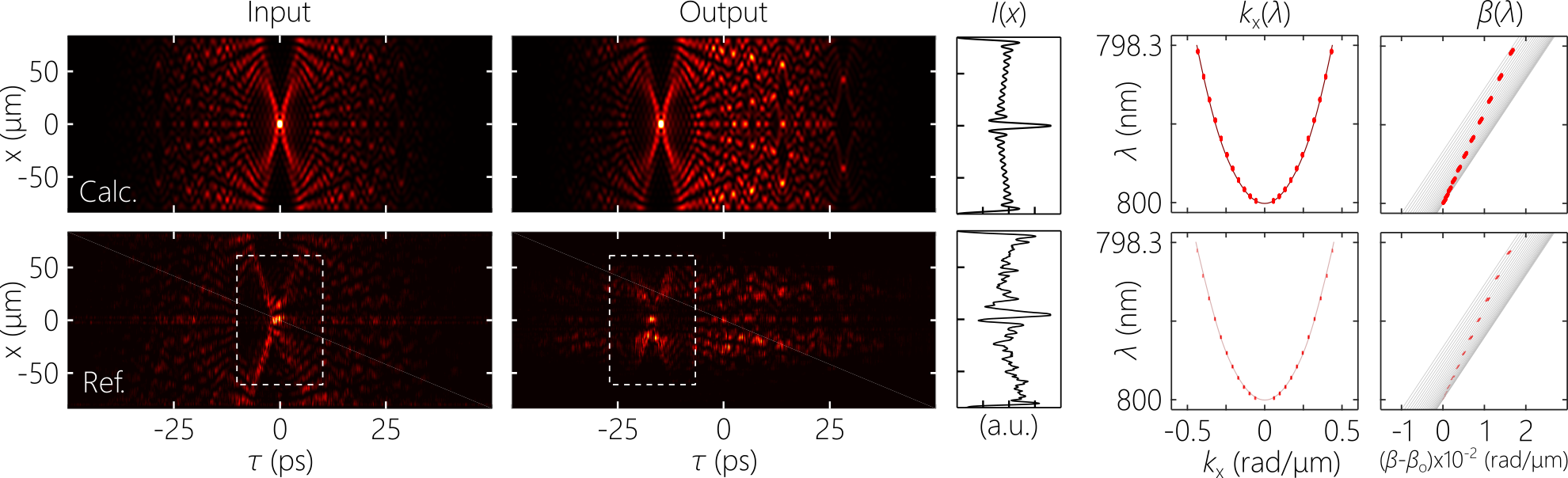}
\caption{\textbf{Superluminal ST supermode.} Columns from left to right correspond to those in Fig.~\ref{Fig:NumberOfModes}. We plot calculations and measurements for a superluminal ST supermode comprising the first 11 even-parity modes with equal weights and $\theta\!=\!44.3^{\circ}$ ($\theta_{\mathrm{a}}\!=\!75^{\circ}$), corresponding to a group velocity $\widetilde{v}\!=\!0.98c\!>\!c/1.51$.}
\label{Fig:Superluminal}
\end{figure*}

\begin{figure}[t!]
\centering
\includegraphics[width=8.5cm]{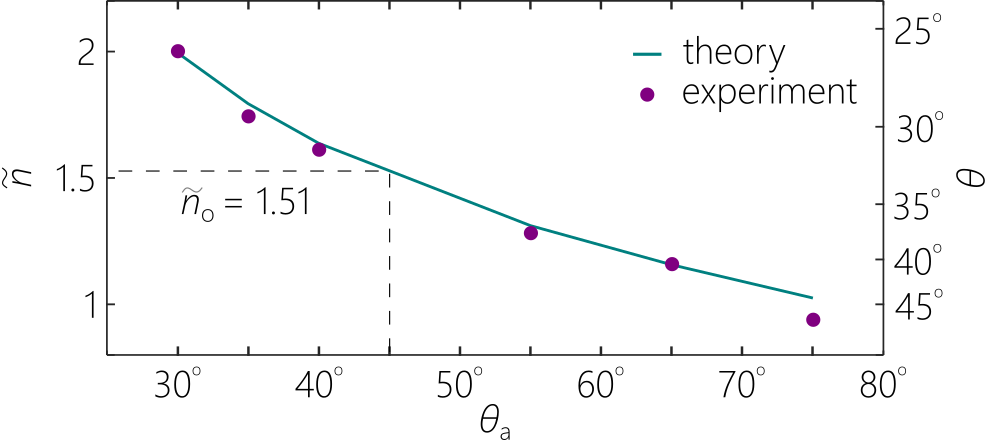}
\caption{\textbf{Tuning the group velocity of ST supermodes.} Measured group index $\widetilde{n}$ for ST supermodes (extracted from the measured group delay with respect to a reference pulse) as we tune the spectral tilt angle $\theta_{\mathrm{a}}$ in free space. The dashed lines identify the luminal condition. The axis on the right shows the corresponding spectral tilt angle $\theta$ of the ST supermode in the waveguide. The solid curve is the theoretical expectation based on Eq.~\ref{Eq:LawOfRefraction}.}
\label{Fig:ExpGroupVelocity}
\end{figure}

We plot in Fig.~\ref{Fig:NumberOfModes}(b-d) the calculated and measured spatio-temporal intensity profiles $I(x;\tau)$ for ST supermodes at the waveguide entrance $z\!=\!0$ and exit $z\!=\!L$, which incorporate an increasing number $M$ of waveguide modes: $M\!=\!4$ [Fig.~\ref{Fig:NumberOfModes}(b)], $M\!=\!7$ [Fig.~\ref{Fig:NumberOfModes}(c)], and $M\!=\!11$ [Fig.~\ref{Fig:NumberOfModes}(d)]. We select here even-parity modes (cosine-like profiles) indexed by $2m+1$, with integer $m\!=\!0,1,2,\cdots$ (the modes corresponding to $m\!=\!0,1$ are eliminated by the spatial filter). Several conclusions can be readily drawn from the measurements. First, the spatial and temporal bandwidths increase with $M$, so that the spatial width at the pulse center decreases from $\Delta x\!\approx\!12$~$\mu$m ($M\!=\!4$) to $\Delta x\!\approx\!6$~$\mu$m ($M\!=\!11$), and the temporal linewidth at the beam center concomitantly decreases from $\Delta T\!\approx\!8$~ps ($M\!=\!4$) to $\Delta T\!\approx\!1.5$~ps ($M\!=\!11$). Consequently, the characteristic X-shaped central feature in the spatio-temporal profile becomes better-defined with increasing $M$. Second, the spatio-temporal profile $I(x;\tau)$ at the exit $z\!=\!L$ resembles that at the entrance $z\!=\!0$ except for two aspects of the X-shaped central feature: (a) it is delayed with respect to the conventional pulsed multimode field (which travels at a group velocity $\approx\!c/n$), thus confirming the subluminal group velocity of the ST supermode; and (b) its position is shifted with respect to the pilot envelope that travels at the same group velocity of the conventional pulsed multimode field. The spectral uncertainty for each mode is $\delta\lambda\!\approx\!42$~pm, and the pilot envelope has a temporal width of $\Delta T\!\approx\!44$~pm. The impact of GVD on these pilot envelopes is thus negligible. Third, the time-averaged intensity $I(x)\!=\!\int\!d\tau \: I(x;\tau)$ measured directly at the waveguide exit by the CCD camera in all cases takes the form of a spatial feature atop an approximately constant pedestal. Fourth, the spatio-temporal spectrum is non-separable with respect to $k_{x}$ and $\lambda$, and takes the form of discrete features (each corresponding to an individual mode) along the parabolic spectrum associated with its freely propagating ST wave packet counterpart having the same group velocity. Finally, the dispersion relationship between $\beta$ and $\lambda$ takes the form of discrete points along a line whose slope corresponds to the group velocity of the ST supermode.

\begin{figure*}[t!]
\centering
\includegraphics[width=15.8cm]{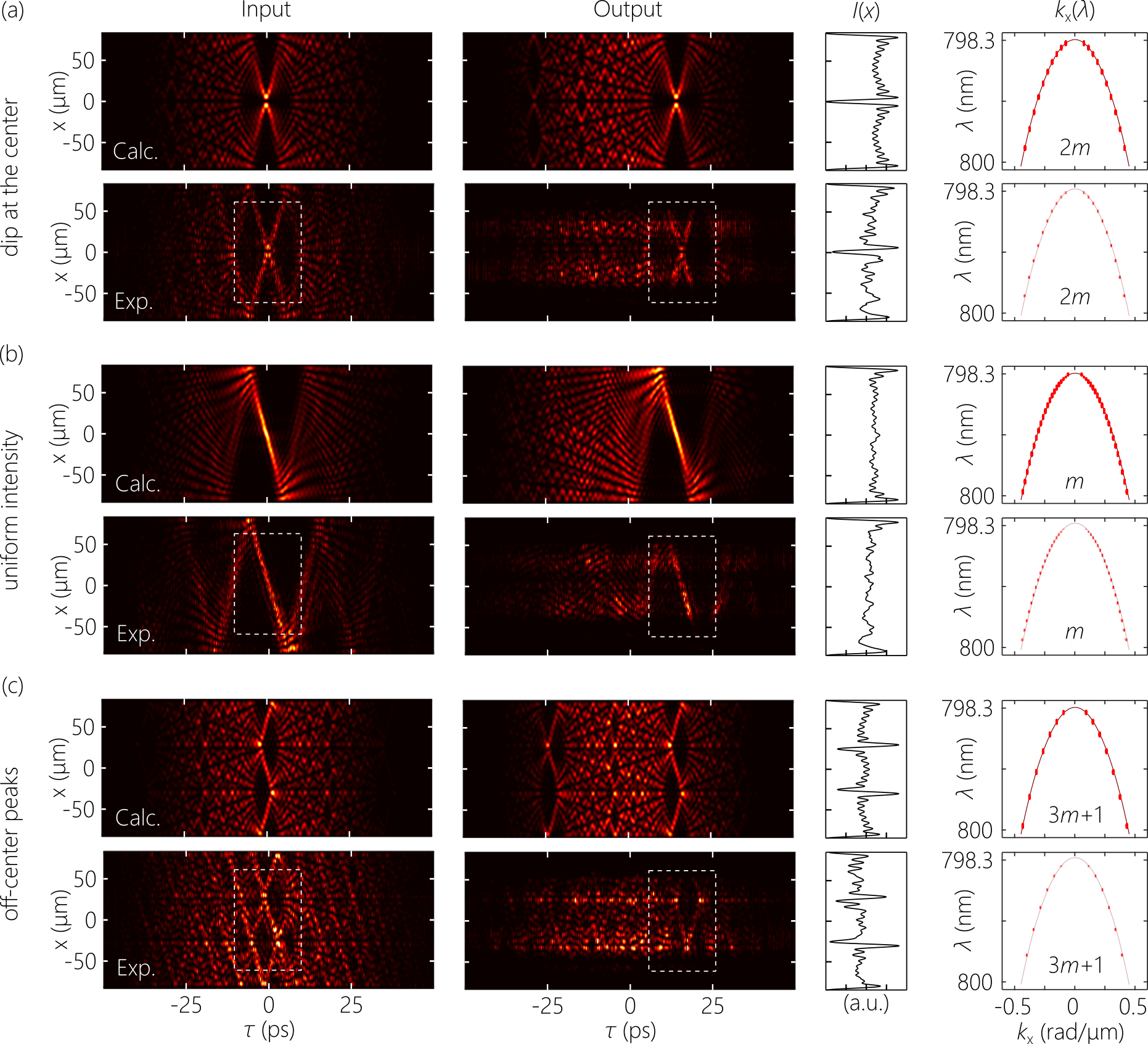}
\caption{\textbf{Modifying the spatio-temporal structure of ST supermodes.} The columns correspond to those in Fig.~\ref{Fig:NumberOfModes}. (a) Selecting 10 odd-parity modes indexed by $2m$ (from $m\!=\!1$ to $m\!=\!10$) produces a dip in the waveguide center. (b) Selecting 21 consecutive modes (comprising alternating even- and odd-parity modes) indexed by $m$ from $m\!=\!1$ to $m\!=\!21$ results in a spatio-temporal profile containing only a single branch from the X-shaped profile, and a spatially uniform time-averaged intensity distribution across the waveguide. (c) Using 6 modes indexed by $3m+1$ (from $m\!=\!1$ to $m\!=\!6$; comprising non-consecutive modes with alternating even- and odd-parity) results in complex spatio-temporal profile containing two peaks, and a time-averaged intensity distribution formed of two peaks atop a constant pedestal. The peaks are separated by $d/3$, and are also at distances $d/3$ from the waveguide walls.}
\label{Fig:ModalShape}
\end{figure*}

\subsection{Tuning the group velocity of ST supermodes}\label{Sec:TuningV}

As shown in Fig.~\ref{Fig:NumberOfModes}(b-d), the X-shaped central feature reaches the waveguide exit at $\tau\!>\!0$, indicating that the ST supermode is slower than the conventional pulsed multimode field [Fig.~\ref{Fig:NumberOfModes}(a)]; here $\tau\!=\!t-nz/c$ is measured in the frame of the conventional pulsed multimode wave packet. If a superluminal ST supermode is synthesized instead, the central peak arrives earlier at $\tau\!<\!0$ than the conventional pulsed field. An example is shown in Fig.~\ref{Fig:Superluminal} for $\widetilde{v}\!=\!0.98c$ ($\theta\!=\!44.3^{\circ}$). All the other features of the ST supermode remain intact, except for the reversal of the sign of curvature of the spatio-temporal spectrum $k_{x}(\lambda)$. Moreover, the group velocity $\widetilde{v}$ of the ST supermodes can be tuned continuously by altering the spectral tilt angle $\theta_{\mathrm{a}}$ of the free-space ST wave packet before coupling to the waveguide. In Fig.~\ref{Fig:ExpGroupVelocity} we plot the estimated ST-supermode group index $\widetilde{n}$ (or spectral tilt angle $\theta$, $\widetilde{n}\!=\!\cot{\theta}$) extracted from the measured group delay with respect to the reference pulse while varying $\theta_{\mathrm{a}}$. In free space, $\theta_{\mathrm{a}}\!=\!45^{\circ}$ ($\widetilde{n}_{\mathrm{a}}\!=\!1$) separates the subluminal and superluminal regimes, whereas $\theta\!\approx\!33.5^{\circ}$ ($\widetilde{n}\!\approx\!n$) separates them in the waveguide. We confirm the tunability of $\widetilde{v}$ for the ST supermode in the waveguide across both the subluminal and superluminal regimes, producing group indices for the ST supermode extending from $\widetilde{n}\!\approx\!1$ to $\widetilde{n}\!\approx\!2$. 

\subsection{Changing the spatial profile of ST supermodes}

The X-shaped spatio-temporal profile of the ST supermodes in Fig.~\ref{Fig:NumberOfModes}(b-d) and the attendant time-averaged intensity peak at the center of the waveguide are a result of choosing only even-parity modes to contribute to the ST supermode. Incidentally, previously demonstrated ST wave packets, whether X-waves or FWMs in free space, have all been X-shaped. However, it has recently been recognized that modulating the spatio-temporal phase can produce propagation-invariant wave packets in free space that are \textit{not} X-shaped \cite{Kondakci18PRL,Wong21OE}. In the context of a ST supermode, modifying the spatio-temporal structure can be realized by careful selection of the waveguide modes that enter into its construction, as demonstrated in Fig.~\ref{Fig:ModalShape}. By selecting odd-parity modes indexed by $2m$, with $m\!=\!1$ to 10, a dip is introduced into the center of the X-shaped spatio-temporal profile and in the time-averaged intensity [Fig.~\ref{Fig:ModalShape}(a)].

Interestingly, when selecting consecutive modes from $m\!=1\!$ to $m\!=\!21$ (comprising alternating even- and odd-parity modes), the X-shaped feature is eliminated and we are left with one branch of that profile [Fig.~\ref{Fig:ModalShape}(b)]. The time-averaged intensity profile is uniform across the waveguide width and remains so with axial propagation. Finally, by selecting modes indexed by $3m+1$ (comprising \textit{non-consecutive} even- and odd-parity modes) from $m\!=\!1$ to $m\!=\!6$ (a total of 6 modes), two off-center peaks are created as shown in Fig.~\ref{Fig:ModalShape}(c). A variety of other structures of ST supermodes may be produced by judicious choice of the modes and their relative weights. 

\section{Discussion}

Although the investigation of free-space ST wave packets extends back to the theoretical breakthrough by Brittingham in 1983 \cite{Brittingham83JAP}, only recently has rapid experimental progress been realized \cite{Yessenov22AOP}. The key to these advances is that ST wave packets can be synthesized with smaller bandwidths and numerical apertures than X-waves or FWMs, while simultaneously offering more versatility with respect to the group-velocity tunability \cite{Yessenov19PRA,Yessenov22AOP}. These features, which make ST wave packets a useful platform for synthesizing ST supermodes, stem from a unique attribute of ST fields: they incorporate \textit{non-differentiable} angular dispersion into their structure \cite{Hall21OL,Hall22OE}. In other words, each frequency $\omega$ propagates at a slightly different angle $\varphi(\omega)$ such that its derivative $\tfrac{d\varphi}{d\omega}$ is \textit{not} defined at some frequency $\omega_{\mathrm{o}}$. It is this non-differentiability that allows for the group velocity of ST wave packets to be tuned across the subluminal and superluminal regimes in the vicinity of $\omega_{\mathrm{o}}$ without recourse to large bandwidths or leaving the paraxial regime \cite{Hall22OE}.

These unique features of ST wave packets have consequently led to increasing interest in \textit{guided} ST fields \cite{Kibler21PRL,Guo21PRR,Bejot21ACSP,Ruano21JO,Bejot22arxiv}. In addition to the ST supermodes reported here, so-called `hybrid ST modes' have been recently synthesized in single-mode \cite{Shiri20NC_Hybrid} and multimode \cite{Shiri21arxiv} planar waveguides. It is important to delineate here these two classes of guided ST fields: ST supermodes and hybrid ST modes in a planar waveguide. In both cases we make use of ST wave packets in the form of a light sheet that is localized along one transverse dimension (say $x$) and extended along the other (say $y$). In the case of a hybrid ST mode, the $x$-axis is aligned with the normal to the waveguide walls. Consequently, the field is localized along one dimension by the confinement mechanism provided by the waveguide, and along the other by the new spatio-temporal structure introduced into the field. The ST wave packet structure in this case can be used to override the usual characteristics of the conventionally guided mode (e.g., eliminate chromatic and modal dispersion, and tune the group index). In contrast, the $x$-axis in the case of ST supermodes is parallel to the planar waveguide walls, and the modulated spatio-temporal structure `interacts' with the interfaces, so that the ST field is constrained to be a superposition of waveguide modes, thereby necessitating spectral discretization. Both of these novel classes of confined fields (hybrid ST modes and ST supermodes) exhibit characteristics that are distinct from those of conventional guided fields.

We have couched our work here in terms of planar-waveguide modes. Can ST supermodes be extended to conventional waveguides in which light is confined in both transverse dimensions? The recent development of so-called 3D ST wave packets that are localized in all dimensions makes this a realizable prospect \cite{Guo21Light,Pang21OL,Yessenov21arx}, especially using the methodology outlined in \cite{Pang21OL} that exploits a spectrally discrete laser comb, whereupon each line in the comb is associated with a particular spatial mode. We therefore expect ST supermodes to be realized in conventional waveguides and fibers in the near future.

Finally, we relate ST supermodes to other recent developments in the physics of optical waveguides. A recent advance has been the realization of `principal modes' in a multimode fiber \cite{Carpenter15NP,Carpenter17LPR}. Waveguide modes are propagation eigenmodes only in absence of scattering. In presence of scattering, new eigenmodes of a selected length of the scattering/dispersive waveguide system can be formed by superposing modes at the same wavelength. Although such a principle mode is \textit{not} invariant along the fiber, it does reproduce the input field distribution at a prescribed output plane. In contrast, ST supermodes remain invariant along the fiber by exploiting the spectral degree of freedom along with the spatial; thus, ST supermodes must be pulsed. However, ST supermodes are \textit{not} immune to scattering. It would therefore be useful to construct `principal ST supermodes' that combine the features of both: propagation invariance \textit{and} immunity to scattering.

Moreover, ST supermodes are related to the recently studied behavior of optical fields interacting with components moving at relativistic speeds \cite{Caloz20IEEE}. Indeed, ST wave packets can be viewed as the result of implementing Lorentz transformation on a monochromatic beam \cite{Belanger86JOSAA,Longhi04OE,Saari04PRE}. For example, a monochromatic laser beam viewed by a detector moving at a velocity $\widetilde{v}$ becomes a propagation-invariant ST wave packet of group velocity $\widetilde{v}$ \cite{Kondakci18PRL}. In this context, a monochromatic field coupled into a waveguide moving at a relativistic velocity induces a multi-chromatic discrete spectral field in the waveguide \cite{Qu16JOSAA}. Our experiments here are thus a confirmation of this result, but with the waveguide stationary while the source is moving with respect to it. This shows the potential utility of ST wave packets in studying interactions between optical fields and moving components in general.

\section{Conclusion}

In conclusion, we have synthesized and observed propagation-invariant ST supermodes in a highly multimoded planar waveguide: a superposition of waveguide modes each assigned to a prescribed wavelength and axial wave number. Precise control over the wavelength and axial wave number assigned to each mode produces an X-shaped spatio-temporal profile that travels rigidly along the multimode waveguide at a prescribed group velocity independently of the group velocity of the waveguide modes. A localized ST supermode at the waveguide input remains localized at its exit, and its time-averaged intensity profile is axially invariant. The propagation-invariance of the spatio-temporal profile is limited by the spectral uncertainty: the finite but small spectral bandwidth assigned to each mode in the ST supermode. By combining up to 21 waveguide modes, we have verified the tunability of the ST-supermode group velocity, and observed a variety of structured intensity profiles, including a dark ST supermode (a dip traveling at the waveguide center), and a flat supermode (a constant intensity profile across the wavelength). These results pave the way to the delivery of pulsed excitations with custom spatial profiles through multimode fibers for linear or nonlinear sensing, imaging, and spectroscopy, and to novel multimode nonlinear interactions in optical waveguides.  

\section*{Funding}
U.S. Office of Naval Research (ONR) contracts N00014-17-1-2458 and N00014-20-1-2789.

\vspace{2mm}
\noindent
\textbf{Disclosures.} The authors declare no conflicts of interest.

\bibliography{diffraction}

\begin{thebibliography}{74}%
\makeatletter
\providecommand \@ifxundefined [1]{%
 \@ifx{#1\undefined}
}%
\providecommand \@ifnum [1]{%
 \ifnum #1\expandafter \@firstoftwo
 \else \expandafter \@secondoftwo
 \fi
}%
\providecommand \@ifx [1]{%
 \ifx #1\expandafter \@firstoftwo
 \else \expandafter \@secondoftwo
 \fi
}%
\providecommand \natexlab [1]{#1}%
\providecommand \enquote  [1]{``#1''}%
\providecommand \bibnamefont  [1]{#1}%
\providecommand \bibfnamefont [1]{#1}%
\providecommand \citenamefont [1]{#1}%
\providecommand \href@noop [0]{\@secondoftwo}%
\providecommand \href [0]{\begingroup \@sanitize@url \@href}%
\providecommand \@href[1]{\@@startlink{#1}\@@href}%
\providecommand \@@href[1]{\endgroup#1\@@endlink}%
\providecommand \@sanitize@url [0]{\catcode `\\12\catcode `\$12\catcode
  `\&12\catcode `\#12\catcode `\^12\catcode `\_12\catcode `\%12\relax}%
\providecommand \@@startlink[1]{}%
\providecommand \@@endlink[0]{}%
\providecommand \url  [0]{\begingroup\@sanitize@url \@url }%
\providecommand \@url [1]{\endgroup\@href {#1}{\urlprefix }}%
\providecommand \urlprefix  [0]{URL }%
\providecommand \Eprint [0]{\href }%
\providecommand \doibase [0]{https://doi.org/}%
\providecommand \selectlanguage [0]{\@gobble}%
\providecommand \bibinfo  [0]{\@secondoftwo}%
\providecommand \bibfield  [0]{\@secondoftwo}%
\providecommand \translation [1]{[#1]}%
\providecommand \BibitemOpen [0]{}%
\providecommand \bibitemStop [0]{}%
\providecommand \bibitemNoStop [0]{.\EOS\space}%
\providecommand \EOS [0]{\spacefactor3000\relax}%
\providecommand \BibitemShut  [1]{\csname bibitem#1\endcsname}%
\let\auto@bib@innerbib\@empty
%</preamble>
\bibitem [{\citenamefont {Mitra}\ and\ \citenamefont {Stark}(2001)}]{Mitra01N}%
  \BibitemOpen
  \bibfield  {author} {\bibinfo {author} {\bibfnamefont {P.~P.}\ \bibnamefont
  {Mitra}}\ and\ \bibinfo {author} {\bibfnamefont {J.~B.}\ \bibnamefont
  {Stark}},\ }\bibfield  {title} {\bibinfo {title} {Nonlinear limits to the
  information capacity of optical fibre communications},\ }\href
  {https://doi.org/10.1038/35082518} {\bibfield  {journal} {\bibinfo  {journal}
  {Nature}\ }\textbf {\bibinfo {volume} {411}},\ \bibinfo {pages} {1027}
  (\bibinfo {year} {2001})}\BibitemShut {NoStop}%
\bibitem [{\citenamefont {Richardson}(2010)}]{Richardson10S}%
  \BibitemOpen
  \bibfield  {author} {\bibinfo {author} {\bibfnamefont {D.~J.}\ \bibnamefont
  {Richardson}},\ }\bibfield  {title} {\bibinfo {title} {Filling the light
  pipe},\ }\href {https://doi.org/10.1126/science.1191708} {\bibfield
  {journal} {\bibinfo  {journal} {Science}\ }\textbf {\bibinfo {volume}
  {330}},\ \bibinfo {pages} {327} (\bibinfo {year} {2010})},\ \Eprint
  {https://arxiv.org/abs/https://www.science.org/doi/pdf/10.1126/science.1191708}
  {https://www.science.org/doi/pdf/10.1126/science.1191708} \BibitemShut
  {NoStop}%
\bibitem [{\citenamefont {Ryf}\ \emph {et~al.}(2013)\citenamefont {Ryf},
  \citenamefont {Randel}, \citenamefont {Fontaine}, \citenamefont {Montoliu},
  \citenamefont {Burrows}, \citenamefont {Corteselli}, \citenamefont
  {Chandrasekhar}, \citenamefont {Gnauck}, \citenamefont {Xie}, \citenamefont
  {Essiambre}, \citenamefont {Winzer}, \citenamefont {Delbue}, \citenamefont
  {Pupalaikis}, \citenamefont {Sureka}, \citenamefont {Sun}, \citenamefont
  {Gr\"{u}ner-Nielsen}, \citenamefont {Jensen},\ and\ \citenamefont
  {Lingle}}]{Ryf13OFC}%
  \BibitemOpen
  \bibfield  {author} {\bibinfo {author} {\bibfnamefont {R.}~\bibnamefont
  {Ryf}}, \bibinfo {author} {\bibfnamefont {S.}~\bibnamefont {Randel}},
  \bibinfo {author} {\bibfnamefont {N.~K.}\ \bibnamefont {Fontaine}}, \bibinfo
  {author} {\bibfnamefont {M.}~\bibnamefont {Montoliu}}, \bibinfo {author}
  {\bibfnamefont {E.}~\bibnamefont {Burrows}}, \bibinfo {author} {\bibfnamefont
  {S.}~\bibnamefont {Corteselli}}, \bibinfo {author} {\bibfnamefont
  {S.}~\bibnamefont {Chandrasekhar}}, \bibinfo {author} {\bibfnamefont {A.~H.}\
  \bibnamefont {Gnauck}}, \bibinfo {author} {\bibfnamefont {C.}~\bibnamefont
  {Xie}}, \bibinfo {author} {\bibfnamefont {R.-J.}\ \bibnamefont {Essiambre}},
  \bibinfo {author} {\bibfnamefont {P.~J.}\ \bibnamefont {Winzer}}, \bibinfo
  {author} {\bibfnamefont {R.}~\bibnamefont {Delbue}}, \bibinfo {author}
  {\bibfnamefont {P.}~\bibnamefont {Pupalaikis}}, \bibinfo {author}
  {\bibfnamefont {A.}~\bibnamefont {Sureka}}, \bibinfo {author} {\bibfnamefont
  {Y.}~\bibnamefont {Sun}}, \bibinfo {author} {\bibfnamefont {L.}~\bibnamefont
  {Gr\"{u}ner-Nielsen}}, \bibinfo {author} {\bibfnamefont {R.~V.}\ \bibnamefont
  {Jensen}},\ and\ \bibinfo {author} {\bibfnamefont {R.}~\bibnamefont
  {Lingle}},\ }\bibfield  {title} {\bibinfo {title} {32-bit/s/{Hz} spectral
  efficiency {WDM} transmission over 177-km few-mode fiber},\ }in\ \href
  {https://doi.org/10.1364/OFC.2013.PDP5A.1} {\emph {\bibinfo {booktitle}
  {Optical Fiber Communication Conference/National Fiber Optic Engineers
  Conference 2013}}}\ (\bibinfo  {publisher} {Optica Publishing Group},\
  \bibinfo {year} {2013})\ p.\ \bibinfo {pages} {PDP5A.1}\BibitemShut {NoStop}%
\bibitem [{\citenamefont {van Uden}\ \emph {et~al.}(2014)\citenamefont {van
  Uden}, \citenamefont {Correa}, \citenamefont {Lopez}, \citenamefont
  {Huijskens}, \citenamefont {Xia}, \citenamefont {Li}, \citenamefont
  {Schülzgen}, \citenamefont {de~Waardt}, \citenamefont {Koonen},\ and\
  \citenamefont {Okonkwo}}]{Uden14NP}%
  \BibitemOpen
  \bibfield  {author} {\bibinfo {author} {\bibfnamefont {R.~G.~H.}\
  \bibnamefont {van Uden}}, \bibinfo {author} {\bibfnamefont {R.~A.}\
  \bibnamefont {Correa}}, \bibinfo {author} {\bibfnamefont {E.~A.}\
  \bibnamefont {Lopez}}, \bibinfo {author} {\bibfnamefont {F.~M.}\ \bibnamefont
  {Huijskens}}, \bibinfo {author} {\bibfnamefont {C.}~\bibnamefont {Xia}},
  \bibinfo {author} {\bibfnamefont {G.}~\bibnamefont {Li}}, \bibinfo {author}
  {\bibfnamefont {A.}~\bibnamefont {Schülzgen}}, \bibinfo {author}
  {\bibfnamefont {H.}~\bibnamefont {de~Waardt}}, \bibinfo {author}
  {\bibfnamefont {A.~M.~J.}\ \bibnamefont {Koonen}},\ and\ \bibinfo {author}
  {\bibfnamefont {C.~M.}\ \bibnamefont {Okonkwo}},\ }\bibfield  {title}
  {\bibinfo {title} {Ultra-high-density spatial division multiplexing with a
  few-mode multicore fibre},\ }\href@noop {} {\bibfield  {journal} {\bibinfo
  {journal} {Nature Photonics}\ }\textbf {\bibinfo {volume} {8}},\ \bibinfo
  {pages} {865} (\bibinfo {year} {2014})}\BibitemShut {NoStop}%
\bibitem [{\citenamefont {Sillard}\ \emph {et~al.}(2016)\citenamefont
  {Sillard}, \citenamefont {Molin}, \citenamefont {Bigot-Astruc}, \citenamefont
  {Jongh}, \citenamefont {Achten}, \citenamefont {Vel\'{a}zquez-Ben\'{i}tez},
  \citenamefont {Amezcua-Correa},\ and\ \citenamefont
  {Okonkwo}}]{Sillard16JLT}%
  \BibitemOpen
  \bibfield  {author} {\bibinfo {author} {\bibfnamefont {P.}~\bibnamefont
  {Sillard}}, \bibinfo {author} {\bibfnamefont {D.}~\bibnamefont {Molin}},
  \bibinfo {author} {\bibfnamefont {M.}~\bibnamefont {Bigot-Astruc}}, \bibinfo
  {author} {\bibfnamefont {K.~D.}\ \bibnamefont {Jongh}}, \bibinfo {author}
  {\bibfnamefont {F.}~\bibnamefont {Achten}}, \bibinfo {author} {\bibfnamefont
  {A.~M.}\ \bibnamefont {Vel\'{a}zquez-Ben\'{i}tez}}, \bibinfo {author}
  {\bibfnamefont {R.}~\bibnamefont {Amezcua-Correa}},\ and\ \bibinfo {author}
  {\bibfnamefont {C.~M.}\ \bibnamefont {Okonkwo}},\ }\bibfield  {title}
  {\bibinfo {title} {Low-differential-mode-group-delay 9-{LP}-mode fiber},\
  }\href@noop {} {\bibfield  {journal} {\bibinfo  {journal} {J. Lightwave
  Technol.}\ }\textbf {\bibinfo {volume} {34}},\ \bibinfo {pages} {425}
  (\bibinfo {year} {2016})}\BibitemShut {NoStop}%
\bibitem [{\citenamefont {Xiong}\ \emph {et~al.}(2018)\citenamefont {Xiong},
  \citenamefont {Hsu}, \citenamefont {Bromberg}, \citenamefont {Antonio-Lopez},
  \citenamefont {Amezcua~Correa},\ and\ \citenamefont {Cao}}]{Xiong18LSA}%
  \BibitemOpen
  \bibfield  {author} {\bibinfo {author} {\bibfnamefont {W.}~\bibnamefont
  {Xiong}}, \bibinfo {author} {\bibfnamefont {C.~W.}\ \bibnamefont {Hsu}},
  \bibinfo {author} {\bibfnamefont {Y.}~\bibnamefont {Bromberg}}, \bibinfo
  {author} {\bibfnamefont {J.~E.}\ \bibnamefont {Antonio-Lopez}}, \bibinfo
  {author} {\bibfnamefont {R.}~\bibnamefont {Amezcua~Correa}},\ and\ \bibinfo
  {author} {\bibfnamefont {H.}~\bibnamefont {Cao}},\ }\bibfield  {title}
  {\bibinfo {title} {Complete polarization control in multimode fibers with
  polarization and mode coupling},\ }\href
  {https://doi.org/10.1038/s41377-018-0047-4} {\bibfield  {journal} {\bibinfo
  {journal} {Light: Science \& Applications}\ }\textbf {\bibinfo {volume}
  {7}},\ \bibinfo {pages} {54} (\bibinfo {year} {2018})}\BibitemShut {NoStop}%
\bibitem [{\citenamefont {Pauwels}\ \emph {et~al.}(2019)\citenamefont
  {Pauwels}, \citenamefont {Van~der Sande},\ and\ \citenamefont
  {Verschaffelt}}]{Pauwels19SR}%
  \BibitemOpen
  \bibfield  {author} {\bibinfo {author} {\bibfnamefont {J.}~\bibnamefont
  {Pauwels}}, \bibinfo {author} {\bibfnamefont {G.}~\bibnamefont {Van~der
  Sande}},\ and\ \bibinfo {author} {\bibfnamefont {G.}~\bibnamefont
  {Verschaffelt}},\ }\bibfield  {title} {\bibinfo {title} {Space division
  multiplexing in standard multi-mode optical fibers based on speckle pattern
  classification},\ }\href@noop {} {\bibfield  {journal} {\bibinfo  {journal}
  {Scientific Reports}\ }\textbf {\bibinfo {volume} {9}},\ \bibinfo {pages}
  {17597} (\bibinfo {year} {2019})}\BibitemShut {NoStop}%
\bibitem [{\citenamefont {Zhou}\ \emph {et~al.}(2021)\citenamefont {Zhou},
  \citenamefont {Braverman}, \citenamefont {Fyffe}, \citenamefont {Zhang},
  \citenamefont {Zhao}, \citenamefont {Willner}, \citenamefont {Shi},\ and\
  \citenamefont {Boyd}}]{Zhou21NC}%
  \BibitemOpen
  \bibfield  {author} {\bibinfo {author} {\bibfnamefont {Y.}~\bibnamefont
  {Zhou}}, \bibinfo {author} {\bibfnamefont {B.}~\bibnamefont {Braverman}},
  \bibinfo {author} {\bibfnamefont {A.}~\bibnamefont {Fyffe}}, \bibinfo
  {author} {\bibfnamefont {R.}~\bibnamefont {Zhang}}, \bibinfo {author}
  {\bibfnamefont {J.}~\bibnamefont {Zhao}}, \bibinfo {author} {\bibfnamefont
  {A.~E.}\ \bibnamefont {Willner}}, \bibinfo {author} {\bibfnamefont
  {Z.}~\bibnamefont {Shi}},\ and\ \bibinfo {author} {\bibfnamefont {R.~W.}\
  \bibnamefont {Boyd}},\ }\bibfield  {title} {\bibinfo {title} {High-fidelity
  spatial mode transmission through a 1-km-long multimode fiber via vectorial
  time reversal},\ }\href {https://doi.org/10.1038/s41467-021-22071-w}
  {\bibfield  {journal} {\bibinfo  {journal} {Nature Communications}\ }\textbf
  {\bibinfo {volume} {12}},\ \bibinfo {pages} {1866} (\bibinfo {year}
  {2021})}\BibitemShut {NoStop}%
\bibitem [{\citenamefont {Zamboni-Rached}\ \emph {et~al.}(2001)\citenamefont
  {Zamboni-Rached}, \citenamefont {Recami},\ and\ \citenamefont
  {Fontana}}]{Zamboni01PRE}%
  \BibitemOpen
  \bibfield  {author} {\bibinfo {author} {\bibfnamefont {M.}~\bibnamefont
  {Zamboni-Rached}}, \bibinfo {author} {\bibfnamefont {E.}~\bibnamefont
  {Recami}},\ and\ \bibinfo {author} {\bibfnamefont {F.}~\bibnamefont
  {Fontana}},\ }\bibfield  {title} {\bibinfo {title} {Superluminal localized
  solutions to {M}axwell equations propagating along a normal-sized
  waveguide},\ }\href@noop {} {\bibfield  {journal} {\bibinfo  {journal} {Phys.
  Rev E}\ }\textbf {\bibinfo {volume} {64}},\ \bibinfo {pages} {066603}
  (\bibinfo {year} {2001})}\BibitemShut {NoStop}%
\bibitem [{\citenamefont {Zamboni-Rached}\ \emph {et~al.}(2002)\citenamefont
  {Zamboni-Rached}, \citenamefont {N{\'o}brega}, \citenamefont {Recami},\ and\
  \citenamefont {Hern{\'a}ndez-Figueroa}}]{Zamboni02PRE}%
  \BibitemOpen
  \bibfield  {author} {\bibinfo {author} {\bibfnamefont {M.}~\bibnamefont
  {Zamboni-Rached}}, \bibinfo {author} {\bibfnamefont {K.~Z.}\ \bibnamefont
  {N{\'o}brega}}, \bibinfo {author} {\bibfnamefont {E.}~\bibnamefont
  {Recami}},\ and\ \bibinfo {author} {\bibfnamefont {H.~E.}\ \bibnamefont
  {Hern{\'a}ndez-Figueroa}},\ }\bibfield  {title} {\bibinfo {title}
  {Superluminal {X}-shaped beams propagating without distortion along a coaxial
  guide},\ }\href@noop {} {\bibfield  {journal} {\bibinfo  {journal} {Phys.
  Rev. E}\ }\textbf {\bibinfo {volume} {66}},\ \bibinfo {pages} {046617}
  (\bibinfo {year} {2002})}\BibitemShut {NoStop}%
\bibitem [{\citenamefont {Zamboni-Rached}\ \emph {et~al.}(2003)\citenamefont
  {Zamboni-Rached}, \citenamefont {Fontana},\ and\ \citenamefont
  {Recami}}]{Zamboni03PRE}%
  \BibitemOpen
  \bibfield  {author} {\bibinfo {author} {\bibfnamefont {M.}~\bibnamefont
  {Zamboni-Rached}}, \bibinfo {author} {\bibfnamefont {F.}~\bibnamefont
  {Fontana}},\ and\ \bibinfo {author} {\bibfnamefont {E.}~\bibnamefont
  {Recami}},\ }\bibfield  {title} {\bibinfo {title} {Superluminal localized
  solutions to {M}axwell equations propagating along a waveguide: the
  finite-energy case},\ }\href@noop {} {\bibfield  {journal} {\bibinfo
  {journal} {Phys. Rev. E}\ }\textbf {\bibinfo {volume} {67}},\ \bibinfo
  {pages} {036620} (\bibinfo {year} {2003})}\BibitemShut {NoStop}%
\bibitem [{\citenamefont {Lu}\ and\ \citenamefont
  {Greenleaf}(1992)}]{Lu92IEEEa}%
  \BibitemOpen
  \bibfield  {author} {\bibinfo {author} {\bibfnamefont {J.-Y.}\ \bibnamefont
  {Lu}}\ and\ \bibinfo {author} {\bibfnamefont {J.~F.}\ \bibnamefont
  {Greenleaf}},\ }\bibfield  {title} {\bibinfo {title} {Nondiffracting {X}
  waves -- exact solutions to free-space scalar wave equation and their finite
  aperture realizations},\ }\href@noop {} {\bibfield  {journal} {\bibinfo
  {journal} {IEEE Trans. Ultrason. Ferroelec. Freq. Control}\ }\textbf
  {\bibinfo {volume} {39}},\ \bibinfo {pages} {19} (\bibinfo {year}
  {1992})}\BibitemShut {NoStop}%
\bibitem [{\citenamefont {Saari}\ and\ \citenamefont
  {Reivelt}(1997)}]{Saari97PRL}%
  \BibitemOpen
  \bibfield  {author} {\bibinfo {author} {\bibfnamefont {P.}~\bibnamefont
  {Saari}}\ and\ \bibinfo {author} {\bibfnamefont {K.}~\bibnamefont
  {Reivelt}},\ }\bibfield  {title} {\bibinfo {title} {Evidence of {X}-shaped
  propagation-invariant localized light waves},\ }\href@noop {} {\bibfield
  {journal} {\bibinfo  {journal} {Phys. Rev. Lett.}\ }\textbf {\bibinfo
  {volume} {79}},\ \bibinfo {pages} {4135} (\bibinfo {year}
  {1997})}\BibitemShut {NoStop}%
\bibitem [{\citenamefont {Reivelt}\ and\ \citenamefont
  {Saari}(2003)}]{Reivelt03arxiv}%
  \BibitemOpen
  \bibfield  {author} {\bibinfo {author} {\bibfnamefont {K.}~\bibnamefont
  {Reivelt}}\ and\ \bibinfo {author} {\bibfnamefont {P.}~\bibnamefont
  {Saari}},\ }\bibfield  {title} {\bibinfo {title} {Localized wave solutions of
  the scalar homogeneous wave equation and their optical implementation},\
  }\href@noop {} {\bibfield  {journal} {\bibinfo  {journal}
  {arxiv:physics/0309079}\ } (\bibinfo {year} {2003})}\BibitemShut {NoStop}%
\bibitem [{\citenamefont {Turunen}\ and\ \citenamefont
  {Friberg}(2010)}]{Turunen10PO}%
  \BibitemOpen
  \bibfield  {author} {\bibinfo {author} {\bibfnamefont {J.}~\bibnamefont
  {Turunen}}\ and\ \bibinfo {author} {\bibfnamefont {A.~T.}\ \bibnamefont
  {Friberg}},\ }\bibfield  {title} {\bibinfo {title} {Propagation-invariant
  optical fields},\ }\href@noop {} {\bibfield  {journal} {\bibinfo  {journal}
  {Prog. Opt.}\ }\textbf {\bibinfo {volume} {54}},\ \bibinfo {pages} {1}
  (\bibinfo {year} {2010})}\BibitemShut {NoStop}%
\bibitem [{\citenamefont {Hern\'andez-Figueroa}\ \emph
  {et~al.}(2014)\citenamefont {Hern\'andez-Figueroa}, \citenamefont {Recami},\
  and\ \citenamefont {Zamboni-Rached}}]{FigueroaBook14}%
  \BibitemOpen
  \bibinfo {editor} {\bibfnamefont {H.~E.}\ \bibnamefont
  {Hern\'andez-Figueroa}}, \bibinfo {editor} {\bibfnamefont {E.}~\bibnamefont
  {Recami}},\ and\ \bibinfo {editor} {\bibfnamefont {M.}~\bibnamefont
  {Zamboni-Rached}},\ eds.,\ \href@noop {} {\emph {\bibinfo {title}
  {Non-diffracting Waves}}}\ (\bibinfo  {publisher} {Wiley-VCH},\ \bibinfo
  {year} {2014})\BibitemShut {NoStop}%
\bibitem [{\citenamefont {Saari}(1996)}]{Saari96Conference}%
  \BibitemOpen
  \bibfield  {author} {\bibinfo {author} {\bibfnamefont {P.}~\bibnamefont
  {Saari}},\ }\bibinfo {title} {Spatially and temporally nondiffracting
  ultrashort pulses},\ in\ \href@noop {} {\emph {\bibinfo {booktitle}
  {Ultrafast Processes in Spectroscopy}}},\ \bibinfo {editor} {edited by\
  \bibinfo {editor} {\bibfnamefont {O.}~\bibnamefont {Svelto}}, \bibinfo
  {editor} {\bibfnamefont {S.}~\bibnamefont {{De S}ilvestri}},\ and\ \bibinfo
  {editor} {\bibfnamefont {G.}~\bibnamefont {Denardo}}}\ (\bibinfo  {publisher}
  {Springer US},\ \bibinfo {address} {Boston, MA},\ \bibinfo {year} {1996})\
  pp.\ \bibinfo {pages} {151--156}\BibitemShut {NoStop}%
\bibitem [{\citenamefont {Grunwald}\ \emph {et~al.}(2003)\citenamefont
  {Grunwald}, \citenamefont {Kebbel}, \citenamefont {Griebner}, \citenamefont
  {Neumann}, \citenamefont {Kummrow}, \citenamefont {Rini}, \citenamefont
  {Nibbering}, \citenamefont {Pich{\'e}}, \citenamefont {Rousseau},\ and\
  \citenamefont {Fortin}}]{Grunwald03PRA}%
  \BibitemOpen
  \bibfield  {author} {\bibinfo {author} {\bibfnamefont {R.}~\bibnamefont
  {Grunwald}}, \bibinfo {author} {\bibfnamefont {V.}~\bibnamefont {Kebbel}},
  \bibinfo {author} {\bibfnamefont {U.}~\bibnamefont {Griebner}}, \bibinfo
  {author} {\bibfnamefont {U.}~\bibnamefont {Neumann}}, \bibinfo {author}
  {\bibfnamefont {A.}~\bibnamefont {Kummrow}}, \bibinfo {author} {\bibfnamefont
  {M.}~\bibnamefont {Rini}}, \bibinfo {author} {\bibfnamefont {E.~T.~J.}\
  \bibnamefont {Nibbering}}, \bibinfo {author} {\bibfnamefont {M.}~\bibnamefont
  {Pich{\'e}}}, \bibinfo {author} {\bibfnamefont {G.}~\bibnamefont
  {Rousseau}},\ and\ \bibinfo {author} {\bibfnamefont {M.}~\bibnamefont
  {Fortin}},\ }\bibfield  {title} {\bibinfo {title} {Generation and
  characterization of spatially and temporally localized few-cycle optical wave
  packets},\ }\href@noop {} {\bibfield  {journal} {\bibinfo  {journal} {Phys.
  Rev. A}\ }\textbf {\bibinfo {volume} {67}},\ \bibinfo {pages} {063820}
  (\bibinfo {year} {2003})}\BibitemShut {NoStop}%
\bibitem [{\citenamefont {Bonaretti}\ \emph {et~al.}(2009)\citenamefont
  {Bonaretti}, \citenamefont {Faccio}, \citenamefont {Clerici}, \citenamefont
  {Biegert},\ and\ \citenamefont {{Di T}rapani}}]{Bonaretti09OE}%
  \BibitemOpen
  \bibfield  {author} {\bibinfo {author} {\bibfnamefont {F.}~\bibnamefont
  {Bonaretti}}, \bibinfo {author} {\bibfnamefont {D.}~\bibnamefont {Faccio}},
  \bibinfo {author} {\bibfnamefont {M.}~\bibnamefont {Clerici}}, \bibinfo
  {author} {\bibfnamefont {J.}~\bibnamefont {Biegert}},\ and\ \bibinfo {author}
  {\bibfnamefont {P.}~\bibnamefont {{Di T}rapani}},\ }\bibfield  {title}
  {\bibinfo {title} {Spatiotemporal amplitude and phase retrieval of
  {B}essel-{X} pulses using a {H}artmann-{S}hack sensor},\ }\href@noop {}
  {\bibfield  {journal} {\bibinfo  {journal} {Opt. Express}\ }\textbf {\bibinfo
  {volume} {17}},\ \bibinfo {pages} {9804} (\bibinfo {year}
  {2009})}\BibitemShut {NoStop}%
\bibitem [{\citenamefont {Bowlan}\ \emph {et~al.}(2009)\citenamefont {Bowlan},
  \citenamefont {Valtna-Lukner}, \citenamefont {L{\~o}hmus}, \citenamefont
  {Piksarv}, \citenamefont {Saari},\ and\ \citenamefont
  {Trebino}}]{Bowlan09OL}%
  \BibitemOpen
  \bibfield  {author} {\bibinfo {author} {\bibfnamefont {P.}~\bibnamefont
  {Bowlan}}, \bibinfo {author} {\bibfnamefont {H.}~\bibnamefont
  {Valtna-Lukner}}, \bibinfo {author} {\bibfnamefont {M.}~\bibnamefont
  {L{\~o}hmus}}, \bibinfo {author} {\bibfnamefont {P.}~\bibnamefont {Piksarv}},
  \bibinfo {author} {\bibfnamefont {P.}~\bibnamefont {Saari}},\ and\ \bibinfo
  {author} {\bibfnamefont {R.}~\bibnamefont {Trebino}},\ }\bibfield  {title}
  {\bibinfo {title} {Measuring the spatiotemporal field of ultrashort
  {B}essel-{X} pulses},\ }\href@noop {} {\bibfield  {journal} {\bibinfo
  {journal} {Opt. Lett.}\ }\textbf {\bibinfo {volume} {34}},\ \bibinfo {pages}
  {2276} (\bibinfo {year} {2009})}\BibitemShut {NoStop}%
\bibitem [{\citenamefont {Kuntz}\ \emph {et~al.}(2009)\citenamefont {Kuntz},
  \citenamefont {Braverman}, \citenamefont {Youn}, \citenamefont {Lobino},
  \citenamefont {Pessina},\ and\ \citenamefont {Lvovsky}}]{Kuntz09PRA}%
  \BibitemOpen
  \bibfield  {author} {\bibinfo {author} {\bibfnamefont {K.~B.}\ \bibnamefont
  {Kuntz}}, \bibinfo {author} {\bibfnamefont {B.}~\bibnamefont {Braverman}},
  \bibinfo {author} {\bibfnamefont {S.~H.}\ \bibnamefont {Youn}}, \bibinfo
  {author} {\bibfnamefont {M.}~\bibnamefont {Lobino}}, \bibinfo {author}
  {\bibfnamefont {E.~M.}\ \bibnamefont {Pessina}},\ and\ \bibinfo {author}
  {\bibfnamefont {A.~I.}\ \bibnamefont {Lvovsky}},\ }\bibfield  {title}
  {\bibinfo {title} {Spatial and temporal characterization of a bessel beam
  produced using a conical mirror},\ }\href@noop {} {\bibfield  {journal}
  {\bibinfo  {journal} {Phys. Rev. A}\ }\textbf {\bibinfo {volume} {79}},\
  \bibinfo {pages} {043802} (\bibinfo {year} {2009})}\BibitemShut {NoStop}%
\bibitem [{\citenamefont {Brittingham}(1983)}]{Brittingham83JAP}%
  \BibitemOpen
  \bibfield  {author} {\bibinfo {author} {\bibfnamefont {J.~N.}\ \bibnamefont
  {Brittingham}},\ }\bibfield  {title} {\bibinfo {title} {Focus wave modes in
  homogeneous {M}axwell's equations: {T}ransverse electric mode},\ }\href@noop
  {} {\bibfield  {journal} {\bibinfo  {journal} {J. Appl. Phys.}\ }\textbf
  {\bibinfo {volume} {54}},\ \bibinfo {pages} {1179} (\bibinfo {year}
  {1983})}\BibitemShut {NoStop}%
\bibitem [{\citenamefont {Vengsarkar}\ \emph {et~al.}(1992)\citenamefont
  {Vengsarkar}, \citenamefont {Besieris}, \citenamefont {Shaarawi},\ and\
  \citenamefont {Ziolkowski}}]{Vengsarkar92JOSAA}%
  \BibitemOpen
  \bibfield  {author} {\bibinfo {author} {\bibfnamefont {A.~M.}\ \bibnamefont
  {Vengsarkar}}, \bibinfo {author} {\bibfnamefont {I.~M.}\ \bibnamefont
  {Besieris}}, \bibinfo {author} {\bibfnamefont {A.~M.}\ \bibnamefont
  {Shaarawi}},\ and\ \bibinfo {author} {\bibfnamefont {R.~W.}\ \bibnamefont
  {Ziolkowski}},\ }\bibfield  {title} {\bibinfo {title} {Closed-form, localized
  wave solutions in optical fiber waveguides},\ }\href@noop {} {\bibfield
  {journal} {\bibinfo  {journal} {J. Opt. Soc. Am. A}\ }\textbf {\bibinfo
  {volume} {9}},\ \bibinfo {pages} {937} (\bibinfo {year} {1992})}\BibitemShut
  {NoStop}%
\bibitem [{\citenamefont {Ruano}\ \emph {et~al.}(2021)\citenamefont {Ruano},
  \citenamefont {Robson},\ and\ \citenamefont {Ornigotti}}]{Ruano21JO}%
  \BibitemOpen
  \bibfield  {author} {\bibinfo {author} {\bibfnamefont {P.~N.}\ \bibnamefont
  {Ruano}}, \bibinfo {author} {\bibfnamefont {C.~W.}\ \bibnamefont {Robson}},\
  and\ \bibinfo {author} {\bibfnamefont {M.}~\bibnamefont {Ornigotti}},\
  }\bibfield  {title} {\bibinfo {title} {Localized waves carrying orbital
  angular momentum in optical fibers},\ }\href@noop {} {\bibfield  {journal}
  {\bibinfo  {journal} {J. Opt.}\ }\textbf {\bibinfo {volume} {23}},\ \bibinfo
  {pages} {075603} (\bibinfo {year} {2021})}\BibitemShut {NoStop}%
\bibitem [{\citenamefont {Yessenov}\ \emph
  {et~al.}(2022{\natexlab{a}})\citenamefont {Yessenov}, \citenamefont {Hall},
  \citenamefont {Schepler},\ and\ \citenamefont {Abouraddy}}]{Yessenov22AOP}%
  \BibitemOpen
  \bibfield  {author} {\bibinfo {author} {\bibfnamefont {M.}~\bibnamefont
  {Yessenov}}, \bibinfo {author} {\bibfnamefont {L.~A.}\ \bibnamefont {Hall}},
  \bibinfo {author} {\bibfnamefont {K.~L.}\ \bibnamefont {Schepler}},\ and\
  \bibinfo {author} {\bibfnamefont {A.~F.}\ \bibnamefont {Abouraddy}},\
  }\bibfield  {title} {\bibinfo {title} {Space-time wave packets},\ }\href@noop
  {} {\bibfield  {journal} {\bibinfo  {journal} {arXiv:2201.08297}\ } (\bibinfo
  {year} {2022}{\natexlab{a}})}\BibitemShut {NoStop}%
\bibitem [{\citenamefont {Kondakci}\ and\ \citenamefont
  {Abouraddy}(2016)}]{Kondakci16OE}%
  \BibitemOpen
  \bibfield  {author} {\bibinfo {author} {\bibfnamefont {H.~E.}\ \bibnamefont
  {Kondakci}}\ and\ \bibinfo {author} {\bibfnamefont {A.~F.}\ \bibnamefont
  {Abouraddy}},\ }\bibfield  {title} {\bibinfo {title} {Diffraction-free pulsed
  optical beams via space-time correlations},\ }\href@noop {} {\bibfield
  {journal} {\bibinfo  {journal} {Opt. Express}\ }\textbf {\bibinfo {volume}
  {24}},\ \bibinfo {pages} {28659} (\bibinfo {year} {2016})}\BibitemShut
  {NoStop}%
\bibitem [{\citenamefont {Parker}\ and\ \citenamefont
  {Alonso}(2016)}]{Parker16OE}%
  \BibitemOpen
  \bibfield  {author} {\bibinfo {author} {\bibfnamefont {K.~J.}\ \bibnamefont
  {Parker}}\ and\ \bibinfo {author} {\bibfnamefont {M.~A.}\ \bibnamefont
  {Alonso}},\ }\bibfield  {title} {\bibinfo {title} {The longitudinal iso-phase
  condition and needle pulses},\ }\href@noop {} {\bibfield  {journal} {\bibinfo
   {journal} {Opt. Express}\ }\textbf {\bibinfo {volume} {24}},\ \bibinfo
  {pages} {28669} (\bibinfo {year} {2016})}\BibitemShut {NoStop}%
\bibitem [{\citenamefont {Yessenov}\ \emph
  {et~al.}(2019{\natexlab{a}})\citenamefont {Yessenov}, \citenamefont
  {Bhaduri}, \citenamefont {Kondakci},\ and\ \citenamefont
  {Abouraddy}}]{Yessenov19OPN}%
  \BibitemOpen
  \bibfield  {author} {\bibinfo {author} {\bibfnamefont {M.}~\bibnamefont
  {Yessenov}}, \bibinfo {author} {\bibfnamefont {B.}~\bibnamefont {Bhaduri}},
  \bibinfo {author} {\bibfnamefont {H.~E.}\ \bibnamefont {Kondakci}},\ and\
  \bibinfo {author} {\bibfnamefont {A.~F.}\ \bibnamefont {Abouraddy}},\
  }\bibfield  {title} {\bibinfo {title} {Weaving the rainbow: Space-time
  optical wave packets},\ }\href@noop {} {\bibfield  {journal} {\bibinfo
  {journal} {Opt. Photon. News}\ }\textbf {\bibinfo {volume} {30}},\ \bibinfo
  {pages} {34} (\bibinfo {year} {2019}{\natexlab{a}})}\BibitemShut {NoStop}%
\bibitem [{\citenamefont {Kondakci}\ and\ \citenamefont
  {Abouraddy}(2017)}]{Kondakci17NP}%
  \BibitemOpen
  \bibfield  {author} {\bibinfo {author} {\bibfnamefont {H.~E.}\ \bibnamefont
  {Kondakci}}\ and\ \bibinfo {author} {\bibfnamefont {A.~F.}\ \bibnamefont
  {Abouraddy}},\ }\bibfield  {title} {\bibinfo {title} {Diffraction-free
  space-time beams},\ }\href@noop {} {\bibfield  {journal} {\bibinfo  {journal}
  {Nat. Photon.}\ }\textbf {\bibinfo {volume} {11}},\ \bibinfo {pages} {733}
  (\bibinfo {year} {2017})}\BibitemShut {NoStop}%
\bibitem [{\citenamefont {Porras}(2017)}]{Porras17OL}%
  \BibitemOpen
  \bibfield  {author} {\bibinfo {author} {\bibfnamefont {M.~A.}\ \bibnamefont
  {Porras}},\ }\bibfield  {title} {\bibinfo {title} {Gaussian beams diffracting
  in time},\ }\href@noop {} {\bibfield  {journal} {\bibinfo  {journal} {Opt.
  Lett.}\ }\textbf {\bibinfo {volume} {42}},\ \bibinfo {pages} {4679} (\bibinfo
  {year} {2017})}\BibitemShut {NoStop}%
\bibitem [{\citenamefont {Efremidis}(2017)}]{Efremidis17OL}%
  \BibitemOpen
  \bibfield  {author} {\bibinfo {author} {\bibfnamefont {N.~K.}\ \bibnamefont
  {Efremidis}},\ }\bibfield  {title} {\bibinfo {title} {Spatiotemporal
  diffraction-free pulsed beams in free-space of the {A}iry and {B}essel
  type},\ }\href@noop {} {\bibfield  {journal} {\bibinfo  {journal} {Opt.
  Lett.}\ }\textbf {\bibinfo {volume} {42}},\ \bibinfo {pages} {5038} (\bibinfo
  {year} {2017})}\BibitemShut {NoStop}%
\bibitem [{\citenamefont {Wong}\ and\ \citenamefont
  {Kaminer}(2017)}]{Wong17ACSP2}%
  \BibitemOpen
  \bibfield  {author} {\bibinfo {author} {\bibfnamefont {L.~J.}\ \bibnamefont
  {Wong}}\ and\ \bibinfo {author} {\bibfnamefont {I.}~\bibnamefont {Kaminer}},\
  }\bibfield  {title} {\bibinfo {title} {Ultrashort tilted-pulsefront pulses
  and nonparaxial tilted-phase-front beams},\ }\href@noop {} {\bibfield
  {journal} {\bibinfo  {journal} {ACS Photon.}\ }\textbf {\bibinfo {volume}
  {4}},\ \bibinfo {pages} {2257} (\bibinfo {year} {2017})}\BibitemShut
  {NoStop}%
\bibitem [{\citenamefont {Kondakci}\ \emph {et~al.}(2018)\citenamefont
  {Kondakci}, \citenamefont {Yessenov}, \citenamefont {Meem}, \citenamefont
  {Reyes}, \citenamefont {Thul}, \citenamefont {Fairchild}, \citenamefont
  {Richardson}, \citenamefont {Menon},\ and\ \citenamefont
  {Abouraddy}}]{Kondakci18OE}%
  \BibitemOpen
  \bibfield  {author} {\bibinfo {author} {\bibfnamefont {H.~E.}\ \bibnamefont
  {Kondakci}}, \bibinfo {author} {\bibfnamefont {M.}~\bibnamefont {Yessenov}},
  \bibinfo {author} {\bibfnamefont {M.}~\bibnamefont {Meem}}, \bibinfo {author}
  {\bibfnamefont {D.}~\bibnamefont {Reyes}}, \bibinfo {author} {\bibfnamefont
  {D.}~\bibnamefont {Thul}}, \bibinfo {author} {\bibfnamefont {S.~R.}\
  \bibnamefont {Fairchild}}, \bibinfo {author} {\bibfnamefont {M.}~\bibnamefont
  {Richardson}}, \bibinfo {author} {\bibfnamefont {R.}~\bibnamefont {Menon}},\
  and\ \bibinfo {author} {\bibfnamefont {A.~F.}\ \bibnamefont {Abouraddy}},\
  }\bibfield  {title} {\bibinfo {title} {Synthesizing broadband
  propagation-invariant space-time wave packets using transmissive phase
  plates},\ }\href@noop {} {\bibfield  {journal} {\bibinfo  {journal} {Opt.
  Express}\ }\textbf {\bibinfo {volume} {26}},\ \bibinfo {pages} {13628}
  (\bibinfo {year} {2018})}\BibitemShut {NoStop}%
\bibitem [{\citenamefont {Wong}\ \emph {et~al.}(2020)\citenamefont {Wong},
  \citenamefont {Christodoulides},\ and\ \citenamefont {Kaminer}}]{Wong20AS}%
  \BibitemOpen
  \bibfield  {author} {\bibinfo {author} {\bibfnamefont {L.~J.}\ \bibnamefont
  {Wong}}, \bibinfo {author} {\bibfnamefont {D.~N.}\ \bibnamefont
  {Christodoulides}},\ and\ \bibinfo {author} {\bibfnamefont {I.}~\bibnamefont
  {Kaminer}},\ }\bibfield  {title} {\bibinfo {title} {The complex charge
  paradigm: {A} new approach for designing electromagnetic wavepackets},\
  }\href@noop {} {\bibfield  {journal} {\bibinfo  {journal} {Adv. Sci.}\
  }\textbf {\bibinfo {volume} {7}},\ \bibinfo {pages} {1903377} (\bibinfo
  {year} {2020})}\BibitemShut {NoStop}%
\bibitem [{\citenamefont {Salo}\ and\ \citenamefont
  {Salomaa}(2001)}]{Salo01JOA}%
  \BibitemOpen
  \bibfield  {author} {\bibinfo {author} {\bibfnamefont {J.}~\bibnamefont
  {Salo}}\ and\ \bibinfo {author} {\bibfnamefont {M.~M.}\ \bibnamefont
  {Salomaa}},\ }\bibfield  {title} {\bibinfo {title} {Diffraction-free pulses
  at arbitrary speeds},\ }\href@noop {} {\bibfield  {journal} {\bibinfo
  {journal} {J. Opt. A}\ }\textbf {\bibinfo {volume} {3}},\ \bibinfo {pages}
  {366} (\bibinfo {year} {2001})}\BibitemShut {NoStop}%
\bibitem [{\citenamefont {Kondakci}\ and\ \citenamefont
  {Abouraddy}(2019)}]{Kondakci19NC}%
  \BibitemOpen
  \bibfield  {author} {\bibinfo {author} {\bibfnamefont {H.~E.}\ \bibnamefont
  {Kondakci}}\ and\ \bibinfo {author} {\bibfnamefont {A.~F.}\ \bibnamefont
  {Abouraddy}},\ }\bibfield  {title} {\bibinfo {title} {Optical space-time wave
  packets of arbitrary group velocity in free space},\ }\href@noop {}
  {\bibfield  {journal} {\bibinfo  {journal} {Nat. Commun.}\ }\textbf {\bibinfo
  {volume} {10}},\ \bibinfo {pages} {929} (\bibinfo {year} {2019})}\BibitemShut
  {NoStop}%
\bibitem [{\citenamefont {Bhaduri}\ \emph {et~al.}(2019)\citenamefont
  {Bhaduri}, \citenamefont {Yessenov},\ and\ \citenamefont
  {Abouraddy}}]{Bhaduri19Optica}%
  \BibitemOpen
  \bibfield  {author} {\bibinfo {author} {\bibfnamefont {B.}~\bibnamefont
  {Bhaduri}}, \bibinfo {author} {\bibfnamefont {M.}~\bibnamefont {Yessenov}},\
  and\ \bibinfo {author} {\bibfnamefont {A.~F.}\ \bibnamefont {Abouraddy}},\
  }\bibfield  {title} {\bibinfo {title} {Space-time wave packets that travel in
  optical materials at the speed of light in vacuum},\ }\href@noop {}
  {\bibfield  {journal} {\bibinfo  {journal} {Optica}\ }\textbf {\bibinfo
  {volume} {6}},\ \bibinfo {pages} {139} (\bibinfo {year} {2019})}\BibitemShut
  {NoStop}%
\bibitem [{\citenamefont {Bhaduri}\ \emph {et~al.}(2020)\citenamefont
  {Bhaduri}, \citenamefont {Yessenov},\ and\ \citenamefont
  {Abouraddy}}]{Bhaduri20NP}%
  \BibitemOpen
  \bibfield  {author} {\bibinfo {author} {\bibfnamefont {B.}~\bibnamefont
  {Bhaduri}}, \bibinfo {author} {\bibfnamefont {M.}~\bibnamefont {Yessenov}},\
  and\ \bibinfo {author} {\bibfnamefont {A.~F.}\ \bibnamefont {Abouraddy}},\
  }\bibfield  {title} {\bibinfo {title} {Anomalous refraction of optical
  spacetime wave packets},\ }\href@noop {} {\bibfield  {journal} {\bibinfo
  {journal} {Nat. Photon.}\ }\textbf {\bibinfo {volume} {14}},\ \bibinfo
  {pages} {416} (\bibinfo {year} {2020})}\BibitemShut {NoStop}%
\bibitem [{\citenamefont {{Allende M}otz}\ \emph {et~al.}(2021)\citenamefont
  {{Allende M}otz}, \citenamefont {Yessenov},\ and\ \citenamefont
  {Abouraddy}}]{AllendeMotz21OL}%
  \BibitemOpen
  \bibfield  {author} {\bibinfo {author} {\bibfnamefont {A.~M.}\ \bibnamefont
  {{Allende M}otz}}, \bibinfo {author} {\bibfnamefont {M.}~\bibnamefont
  {Yessenov}},\ and\ \bibinfo {author} {\bibfnamefont {A.~F.}\ \bibnamefont
  {Abouraddy}},\ }\bibfield  {title} {\bibinfo {title} {Isochronous space-time
  wave packets},\ }\href@noop {} {\bibfield  {journal} {\bibinfo  {journal}
  {Opt. Lett.}\ }\textbf {\bibinfo {volume} {46}},\ \bibinfo {pages} {2260}
  (\bibinfo {year} {2021})}\BibitemShut {NoStop}%
\bibitem [{\citenamefont {Hall}\ \emph
  {et~al.}(2021{\natexlab{a}})\citenamefont {Hall}, \citenamefont {Yessenov},
  \citenamefont {Ponomarenko},\ and\ \citenamefont {Abouraddy}}]{Hall21APL}%
  \BibitemOpen
  \bibfield  {author} {\bibinfo {author} {\bibfnamefont {L.~A.}\ \bibnamefont
  {Hall}}, \bibinfo {author} {\bibfnamefont {M.}~\bibnamefont {Yessenov}},
  \bibinfo {author} {\bibfnamefont {S.~A.}\ \bibnamefont {Ponomarenko}},\ and\
  \bibinfo {author} {\bibfnamefont {A.~F.}\ \bibnamefont {Abouraddy}},\
  }\bibfield  {title} {\bibinfo {title} {The space-time {T}albot effect},\
  }\href@noop {} {\bibfield  {journal} {\bibinfo  {journal} {APL Photon.}\
  }\textbf {\bibinfo {volume} {6}},\ \bibinfo {pages} {056105} (\bibinfo {year}
  {2021}{\natexlab{a}})}\BibitemShut {NoStop}%
\bibitem [{\citenamefont {Orlov}\ \emph {et~al.}(2002)\citenamefont {Orlov},
  \citenamefont {Piskarskas},\ and\ \citenamefont {Stabinis}}]{Orlov02OL}%
  \BibitemOpen
  \bibfield  {author} {\bibinfo {author} {\bibfnamefont {S.}~\bibnamefont
  {Orlov}}, \bibinfo {author} {\bibfnamefont {A.}~\bibnamefont {Piskarskas}},\
  and\ \bibinfo {author} {\bibfnamefont {A.}~\bibnamefont {Stabinis}},\
  }\bibfield  {title} {\bibinfo {title} {Localized optical subcycle pulses in
  dispersive media},\ }\href@noop {} {\bibfield  {journal} {\bibinfo  {journal}
  {Opt. Lett.}\ }\textbf {\bibinfo {volume} {27}},\ \bibinfo {pages} {2167}
  (\bibinfo {year} {2002})}\BibitemShut {NoStop}%
\bibitem [{\citenamefont {Porras}\ \emph {et~al.}(2003)\citenamefont {Porras},
  \citenamefont {Valiulis},\ and\ \citenamefont {{Di T}rapani}}]{Porras03PRE2}%
  \BibitemOpen
  \bibfield  {author} {\bibinfo {author} {\bibfnamefont {M.~A.}\ \bibnamefont
  {Porras}}, \bibinfo {author} {\bibfnamefont {G.}~\bibnamefont {Valiulis}},\
  and\ \bibinfo {author} {\bibfnamefont {P.}~\bibnamefont {{Di T}rapani}},\
  }\bibfield  {title} {\bibinfo {title} {Unified description of {B}essel {X}
  waves with cone dispersion and tilted pulses},\ }\href@noop {} {\bibfield
  {journal} {\bibinfo  {journal} {Phys. Rev. E}\ }\textbf {\bibinfo {volume}
  {68}},\ \bibinfo {pages} {016613} (\bibinfo {year} {2003})}\BibitemShut
  {NoStop}%
\bibitem [{\citenamefont {Longhi}(2004{\natexlab{a}})}]{Longhi04OL}%
  \BibitemOpen
  \bibfield  {author} {\bibinfo {author} {\bibfnamefont {S.}~\bibnamefont
  {Longhi}},\ }\bibfield  {title} {\bibinfo {title} {Localized subluminal
  envelope pulses in dispersive media},\ }\href@noop {} {\bibfield  {journal}
  {\bibinfo  {journal} {Opt. Lett.}\ }\textbf {\bibinfo {volume} {29}},\
  \bibinfo {pages} {147} (\bibinfo {year} {2004}{\natexlab{a}})}\BibitemShut
  {NoStop}%
\bibitem [{\citenamefont {Porras}\ and\ \citenamefont {{Di
  T}rapani}(2004)}]{Porras04PRE}%
  \BibitemOpen
  \bibfield  {author} {\bibinfo {author} {\bibfnamefont {M.~A.}\ \bibnamefont
  {Porras}}\ and\ \bibinfo {author} {\bibfnamefont {P.}~\bibnamefont {{Di
  T}rapani}},\ }\bibfield  {title} {\bibinfo {title} {Localized and stationary
  light wave modes in dispersive media},\ }\href@noop {} {\bibfield  {journal}
  {\bibinfo  {journal} {Phys. Rev. E}\ }\textbf {\bibinfo {volume} {69}},\
  \bibinfo {pages} {066606} (\bibinfo {year} {2004})}\BibitemShut {NoStop}%
\bibitem [{\citenamefont {Malaguti}\ \emph {et~al.}(2008)\citenamefont
  {Malaguti}, \citenamefont {Bellanca},\ and\ \citenamefont
  {Trillo}}]{Malaguti08OL}%
  \BibitemOpen
  \bibfield  {author} {\bibinfo {author} {\bibfnamefont {S.}~\bibnamefont
  {Malaguti}}, \bibinfo {author} {\bibfnamefont {G.}~\bibnamefont {Bellanca}},\
  and\ \bibinfo {author} {\bibfnamefont {S.}~\bibnamefont {Trillo}},\
  }\bibfield  {title} {\bibinfo {title} {Two-dimensional envelope localized
  waves in the anomalous dispersion regime},\ }\href@noop {} {\bibfield
  {journal} {\bibinfo  {journal} {Opt. Lett.}\ }\textbf {\bibinfo {volume}
  {33}},\ \bibinfo {pages} {1117} (\bibinfo {year} {2008})}\BibitemShut
  {NoStop}%
\bibitem [{\citenamefont {Malaguti}\ and\ \citenamefont
  {Trillo}(2009)}]{Malaguti09PRA}%
  \BibitemOpen
  \bibfield  {author} {\bibinfo {author} {\bibfnamefont {S.}~\bibnamefont
  {Malaguti}}\ and\ \bibinfo {author} {\bibfnamefont {S.}~\bibnamefont
  {Trillo}},\ }\bibfield  {title} {\bibinfo {title} {Envelope localized waves
  of the conical type in linear normally dispersive media},\ }\href@noop {}
  {\bibfield  {journal} {\bibinfo  {journal} {Phys. Rev. A}\ }\textbf {\bibinfo
  {volume} {79}},\ \bibinfo {pages} {063803} (\bibinfo {year}
  {2009})}\BibitemShut {NoStop}%
\bibitem [{\citenamefont {Yessenov}\ \emph
  {et~al.}(2021{\natexlab{a}})\citenamefont {Yessenov}, \citenamefont {Hall},\
  and\ \citenamefont {Abouraddy}}]{Yessenov21ACSPhot}%
  \BibitemOpen
  \bibfield  {author} {\bibinfo {author} {\bibfnamefont {M.}~\bibnamefont
  {Yessenov}}, \bibinfo {author} {\bibfnamefont {L.~A.}\ \bibnamefont {Hall}},\
  and\ \bibinfo {author} {\bibfnamefont {A.~F.}\ \bibnamefont {Abouraddy}},\
  }\bibfield  {title} {\bibinfo {title} {Engineering the optical vacuum:
  Arbitrary magnitude, sign, and order of dispersion in free space using
  space–time wave packets},\ }\href@noop {} {\bibfield  {journal} {\bibinfo
  {journal} {ACS Photon.}\ }\textbf {\bibinfo {volume} {8}},\ \bibinfo {pages}
  {2274} (\bibinfo {year} {2021}{\natexlab{a}})}\BibitemShut {NoStop}%
\bibitem [{\citenamefont {Hall}\ and\ \citenamefont
  {Abouraddy}(2022{\natexlab{a}})}]{Hall22LPR}%
  \BibitemOpen
  \bibfield  {author} {\bibinfo {author} {\bibfnamefont {L.~A.}\ \bibnamefont
  {Hall}}\ and\ \bibinfo {author} {\bibfnamefont {A.~F.}\ \bibnamefont
  {Abouraddy}},\ }\bibfield  {title} {\bibinfo {title} {Canceling and inverting
  normal and anomalous group-velocity dispersion using space-time wave
  packets},\ }\href@noop {} {\bibfield  {journal} {\bibinfo  {journal}
  {arXiv:2202.01148}\ } (\bibinfo {year} {2022}{\natexlab{a}})}\BibitemShut
  {NoStop}%
\bibitem [{\citenamefont {Guo}\ and\ \citenamefont {Fan}(2021)}]{Guo21PRR}%
  \BibitemOpen
  \bibfield  {author} {\bibinfo {author} {\bibfnamefont {C.}~\bibnamefont
  {Guo}}\ and\ \bibinfo {author} {\bibfnamefont {S.}~\bibnamefont {Fan}},\
  }\bibfield  {title} {\bibinfo {title} {Generation of guided space-time wave
  packets using multilevel indirect photonic transitions in integrated
  photonics},\ }\href@noop {} {\bibfield  {journal} {\bibinfo  {journal} {Phys.
  Rev. Research}\ }\textbf {\bibinfo {volume} {3}},\ \bibinfo {pages} {033161}
  (\bibinfo {year} {2021})}\BibitemShut {NoStop}%
\bibitem [{\citenamefont {Kibler}\ and\ \citenamefont
  {B{\'e}jot}(2021)}]{Kibler21PRL}%
  \BibitemOpen
  \bibfield  {author} {\bibinfo {author} {\bibfnamefont {B.}~\bibnamefont
  {Kibler}}\ and\ \bibinfo {author} {\bibfnamefont {P.}~\bibnamefont
  {B{\'e}jot}},\ }\bibfield  {title} {\bibinfo {title} {Discretized conical
  waves in multimode optical fibers},\ }\href@noop {} {\bibfield  {journal}
  {\bibinfo  {journal} {Phys. Rev. Lett.}\ }\textbf {\bibinfo {volume} {126}},\
  \bibinfo {pages} {023902} (\bibinfo {year} {2021})}\BibitemShut {NoStop}%
\bibitem [{\citenamefont {B{\'e}jot}\ and\ \citenamefont
  {Kibler}(2021)}]{Bejot21ACSP}%
  \BibitemOpen
  \bibfield  {author} {\bibinfo {author} {\bibfnamefont {P.}~\bibnamefont
  {B{\'e}jot}}\ and\ \bibinfo {author} {\bibfnamefont {B.}~\bibnamefont
  {Kibler}},\ }\bibfield  {title} {\bibinfo {title} {Spatiotemporal helicon
  wavepackets},\ }\href@noop {} {\bibfield  {journal} {\bibinfo  {journal} {ACS
  Photon.}\ }\textbf {\bibinfo {volume} {8}},\ \bibinfo {pages} {2345}
  (\bibinfo {year} {2021})}\BibitemShut {NoStop}%
\bibitem [{\citenamefont {Donnelly}\ and\ \citenamefont
  {Ziolkowski}(1993)}]{Donnelly93ProcRSLA}%
  \BibitemOpen
  \bibfield  {author} {\bibinfo {author} {\bibfnamefont {R.}~\bibnamefont
  {Donnelly}}\ and\ \bibinfo {author} {\bibfnamefont {R.~W.}\ \bibnamefont
  {Ziolkowski}},\ }\bibfield  {title} {\bibinfo {title} {Designing localized
  waves},\ }\href@noop {} {\bibfield  {journal} {\bibinfo  {journal} {Proc. R.
  Soc. Lond. A}\ }\textbf {\bibinfo {volume} {440}},\ \bibinfo {pages} {541}
  (\bibinfo {year} {1993})}\BibitemShut {NoStop}%
\bibitem [{\citenamefont {Yessenov}\ \emph
  {et~al.}(2019{\natexlab{b}})\citenamefont {Yessenov}, \citenamefont
  {Bhaduri}, \citenamefont {Kondakci},\ and\ \citenamefont
  {Abouraddy}}]{Yessenov19PRA}%
  \BibitemOpen
  \bibfield  {author} {\bibinfo {author} {\bibfnamefont {M.}~\bibnamefont
  {Yessenov}}, \bibinfo {author} {\bibfnamefont {B.}~\bibnamefont {Bhaduri}},
  \bibinfo {author} {\bibfnamefont {H.~E.}\ \bibnamefont {Kondakci}},\ and\
  \bibinfo {author} {\bibfnamefont {A.~F.}\ \bibnamefont {Abouraddy}},\
  }\bibfield  {title} {\bibinfo {title} {Classification of
  propagation-invariant space-time light-sheets in free space: Theory and
  experiments},\ }\href@noop {} {\bibfield  {journal} {\bibinfo  {journal}
  {Phys. Rev. A}\ }\textbf {\bibinfo {volume} {99}},\ \bibinfo {pages} {023856}
  (\bibinfo {year} {2019}{\natexlab{b}})}\BibitemShut {NoStop}%
\bibitem [{\citenamefont {Kondakci}\ and\ \citenamefont
  {Abouraddy}(2018)}]{Kondakci18PRL}%
  \BibitemOpen
  \bibfield  {author} {\bibinfo {author} {\bibfnamefont {H.~E.}\ \bibnamefont
  {Kondakci}}\ and\ \bibinfo {author} {\bibfnamefont {A.~F.}\ \bibnamefont
  {Abouraddy}},\ }\bibfield  {title} {\bibinfo {title} {Airy wavepackets
  accelerating in space-time},\ }\href@noop {} {\bibfield  {journal} {\bibinfo
  {journal} {Phys. Rev. Lett.}\ }\textbf {\bibinfo {volume} {120}},\ \bibinfo
  {pages} {163901} (\bibinfo {year} {2018})}\BibitemShut {NoStop}%
\bibitem [{\citenamefont {Wong}(2021)}]{Wong21OE}%
  \BibitemOpen
  \bibfield  {author} {\bibinfo {author} {\bibfnamefont {L.~J.}\ \bibnamefont
  {Wong}},\ }\bibfield  {title} {\bibinfo {title} {Propagation-invariant
  space-time caustics of light},\ }\href@noop {} {\bibfield  {journal}
  {\bibinfo  {journal} {Opt. Express}\ }\textbf {\bibinfo {volume} {29}},\
  \bibinfo {pages} {30682} (\bibinfo {year} {2021})}\BibitemShut {NoStop}%
\bibitem [{\citenamefont {Sezginer}(1985)}]{Sezginer85JAP}%
  \BibitemOpen
  \bibfield  {author} {\bibinfo {author} {\bibfnamefont {A.}~\bibnamefont
  {Sezginer}},\ }\bibfield  {title} {\bibinfo {title} {A general formulation of
  focus wave modes},\ }\href@noop {} {\bibfield  {journal} {\bibinfo  {journal}
  {J. Appl. Phys.}\ }\textbf {\bibinfo {volume} {57}},\ \bibinfo {pages} {678}
  (\bibinfo {year} {1985})}\BibitemShut {NoStop}%
\bibitem [{\citenamefont {Yessenov}\ \emph
  {et~al.}(2019{\natexlab{c}})\citenamefont {Yessenov}, \citenamefont
  {Bhaduri}, \citenamefont {Mach}, \citenamefont {Mardani}, \citenamefont
  {Kondakci}, \citenamefont {Alonso}, \citenamefont {Atia},\ and\ \citenamefont
  {Abouraddy}}]{Yessenov19OE}%
  \BibitemOpen
  \bibfield  {author} {\bibinfo {author} {\bibfnamefont {M.}~\bibnamefont
  {Yessenov}}, \bibinfo {author} {\bibfnamefont {B.}~\bibnamefont {Bhaduri}},
  \bibinfo {author} {\bibfnamefont {L.}~\bibnamefont {Mach}}, \bibinfo {author}
  {\bibfnamefont {D.}~\bibnamefont {Mardani}}, \bibinfo {author} {\bibfnamefont
  {H.~E.}\ \bibnamefont {Kondakci}}, \bibinfo {author} {\bibfnamefont {M.~A.}\
  \bibnamefont {Alonso}}, \bibinfo {author} {\bibfnamefont {G.~A.}\
  \bibnamefont {Atia}},\ and\ \bibinfo {author} {\bibfnamefont {A.~F.}\
  \bibnamefont {Abouraddy}},\ }\bibfield  {title} {\bibinfo {title} {What is
  the maximum differential group delay achievable by a space-time wave packet
  in free space?},\ }\href@noop {} {\bibfield  {journal} {\bibinfo  {journal}
  {Opt. Express}\ }\textbf {\bibinfo {volume} {27}},\ \bibinfo {pages} {12443}
  (\bibinfo {year} {2019}{\natexlab{c}})}\BibitemShut {NoStop}%
\bibitem [{\citenamefont {He}\ \emph {et~al.}(2021)\citenamefont {He},
  \citenamefont {Guo},\ and\ \citenamefont {Xiao}}]{He21arxiv}%
  \BibitemOpen
  \bibfield  {author} {\bibinfo {author} {\bibfnamefont {H.}~\bibnamefont
  {He}}, \bibinfo {author} {\bibfnamefont {C.}~\bibnamefont {Guo}},\ and\
  \bibinfo {author} {\bibfnamefont {M.}~\bibnamefont {Xiao}},\ }\bibfield
  {title} {\bibinfo {title} {Non-dispersive space-time wave packets propagating
  in dispersive media},\ }\href@noop {} {\bibfield  {journal} {\bibinfo
  {journal} {arXiv:2109.00782}\ } (\bibinfo {year} {2021})}\BibitemShut
  {NoStop}%
\bibitem [{\citenamefont {Yessenov}\ \emph
  {et~al.}(2022{\natexlab{b}})\citenamefont {Yessenov}, \citenamefont
  {Faryadras}, \citenamefont {Benis}, \citenamefont {Hagan}, \citenamefont {{E.
  W. Van S}tryland},\ and\ \citenamefont {Abouraddy}}]{Yessenov22OL}%
  \BibitemOpen
  \bibfield  {author} {\bibinfo {author} {\bibfnamefont {M.}~\bibnamefont
  {Yessenov}}, \bibinfo {author} {\bibfnamefont {S.}~\bibnamefont {Faryadras}},
  \bibinfo {author} {\bibfnamefont {S.}~\bibnamefont {Benis}}, \bibinfo
  {author} {\bibfnamefont {D.~J.}\ \bibnamefont {Hagan}}, \bibinfo {author}
  {\bibnamefont {{E. W. Van S}tryland}},\ and\ \bibinfo {author} {\bibfnamefont
  {A.~F.}\ \bibnamefont {Abouraddy}},\ }\bibfield  {title} {\bibinfo {title}
  {Refraction of space–time wave packets in a dispersive medium},\
  }\href@noop {} {\bibfield  {journal} {\bibinfo  {journal} {Opt. Lett.}\
  }\textbf {\bibinfo {volume} {47}},\ \bibinfo {pages} {1630} (\bibinfo {year}
  {2022}{\natexlab{b}})}\BibitemShut {NoStop}%
\bibitem [{\citenamefont {Hall}\ \emph
  {et~al.}(2021{\natexlab{b}})\citenamefont {Hall}, \citenamefont {Yessenov},\
  and\ \citenamefont {Abouraddy}}]{Hall21OL}%
  \BibitemOpen
  \bibfield  {author} {\bibinfo {author} {\bibfnamefont {L.~A.}\ \bibnamefont
  {Hall}}, \bibinfo {author} {\bibfnamefont {M.}~\bibnamefont {Yessenov}},\
  and\ \bibinfo {author} {\bibfnamefont {A.~F.}\ \bibnamefont {Abouraddy}},\
  }\bibfield  {title} {\bibinfo {title} {Space--time wave packets violate the
  universal relationship between angular dispersion and pulse-front tilt},\
  }\href@noop {} {\bibfield  {journal} {\bibinfo  {journal} {Opt. Lett.}\
  }\textbf {\bibinfo {volume} {46}},\ \bibinfo {pages} {1672} (\bibinfo {year}
  {2021}{\natexlab{b}})}\BibitemShut {NoStop}%
\bibitem [{\citenamefont {Hall}\ and\ \citenamefont
  {Abouraddy}(2022{\natexlab{b}})}]{Hall22OE}%
  \BibitemOpen
  \bibfield  {author} {\bibinfo {author} {\bibfnamefont {L.~A.}\ \bibnamefont
  {Hall}}\ and\ \bibinfo {author} {\bibfnamefont {A.~F.}\ \bibnamefont
  {Abouraddy}},\ }\bibfield  {title} {\bibinfo {title} {Consequences of
  non-differentiable angular dispersion in optics: {T}ilted pulse fronts versus
  space-time wave packets},\ }\href@noop {} {\bibfield  {journal} {\bibinfo
  {journal} {Opt. Express}\ }\textbf {\bibinfo {volume} {30}},\ \bibinfo
  {pages} {4817} (\bibinfo {year} {2022}{\natexlab{b}})}\BibitemShut {NoStop}%
\bibitem [{\citenamefont {B{\'e}jot}\ and\ \citenamefont
  {Kibler}(2022)}]{Bejot22arxiv}%
  \BibitemOpen
  \bibfield  {author} {\bibinfo {author} {\bibfnamefont {P.}~\bibnamefont
  {B{\'e}jot}}\ and\ \bibinfo {author} {\bibfnamefont {B.}~\bibnamefont
  {Kibler}},\ }\bibfield  {title} {\bibinfo {title} {Quadrics for structuring
  space-time wavepackets},\ }\href@noop {} {\bibfield  {journal} {\bibinfo
  {journal} {arXiv:2202.00407}\ } (\bibinfo {year} {2022})}\BibitemShut
  {NoStop}%
\bibitem [{\citenamefont {Shiri}\ \emph {et~al.}(2020)\citenamefont {Shiri},
  \citenamefont {Yessenov}, \citenamefont {Webster}, \citenamefont {Schepler},\
  and\ \citenamefont {Abouraddy}}]{Shiri20NC_Hybrid}%
  \BibitemOpen
  \bibfield  {author} {\bibinfo {author} {\bibfnamefont {A.}~\bibnamefont
  {Shiri}}, \bibinfo {author} {\bibfnamefont {M.}~\bibnamefont {Yessenov}},
  \bibinfo {author} {\bibfnamefont {S.}~\bibnamefont {Webster}}, \bibinfo
  {author} {\bibfnamefont {K.~L.}\ \bibnamefont {Schepler}},\ and\ \bibinfo
  {author} {\bibfnamefont {A.~F.}\ \bibnamefont {Abouraddy}},\ }\bibfield
  {title} {\bibinfo {title} {Hybrid guided space-time optical modes in
  unpatterned films},\ }\href@noop {} {\bibfield  {journal} {\bibinfo
  {journal} {Nat. Commun.}\ }\textbf {\bibinfo {volume} {11}},\ \bibinfo
  {pages} {6273} (\bibinfo {year} {2020})}\BibitemShut {NoStop}%
\bibitem [{\citenamefont {Shiri}\ and\ \citenamefont
  {Abouraddy}(2021)}]{Shiri21arxiv}%
  \BibitemOpen
  \bibfield  {author} {\bibinfo {author} {\bibfnamefont {A.}~\bibnamefont
  {Shiri}}\ and\ \bibinfo {author} {\bibfnamefont {A.~F.}\ \bibnamefont
  {Abouraddy}},\ }\bibfield  {title} {\bibinfo {title} {Severing the link
  between modal order and group index using hybrid guided space-time modes},\
  }\href@noop {} {\bibfield  {journal} {\bibinfo  {journal} {arXiv:2111.02617}\
  } (\bibinfo {year} {2021})}\BibitemShut {NoStop}%
\bibitem [{\citenamefont {Guo}\ \emph {et~al.}(2021)\citenamefont {Guo},
  \citenamefont {Xiao}, \citenamefont {Orenstein},\ and\ \citenamefont
  {Fan}}]{Guo21Light}%
  \BibitemOpen
  \bibfield  {author} {\bibinfo {author} {\bibfnamefont {C.}~\bibnamefont
  {Guo}}, \bibinfo {author} {\bibfnamefont {M.}~\bibnamefont {Xiao}}, \bibinfo
  {author} {\bibfnamefont {M.}~\bibnamefont {Orenstein}},\ and\ \bibinfo
  {author} {\bibfnamefont {S.}~\bibnamefont {Fan}},\ }\bibfield  {title}
  {\bibinfo {title} {Structured {3D} linear space--time light bullets by
  nonlocal nanophotonics},\ }\href@noop {} {\bibfield  {journal} {\bibinfo
  {journal} {Light: Sci. Appl.}\ }\textbf {\bibinfo {volume} {10}},\ \bibinfo
  {pages} {160} (\bibinfo {year} {2021})}\BibitemShut {NoStop}%
\bibitem [{\citenamefont {Pang}\ \emph {et~al.}(2021)\citenamefont {Pang},
  \citenamefont {Zou}, \citenamefont {Song}, \citenamefont {Zhao},
  \citenamefont {Minoofar}, \citenamefont {Zhang}, \citenamefont {Song},
  \citenamefont {Zhou}, \citenamefont {Su}, \citenamefont {Liu}, \citenamefont
  {Hu}, \citenamefont {Tur},\ and\ \citenamefont {Willner}}]{Pang21OL}%
  \BibitemOpen
  \bibfield  {author} {\bibinfo {author} {\bibfnamefont {K.}~\bibnamefont
  {Pang}}, \bibinfo {author} {\bibfnamefont {K.}~\bibnamefont {Zou}}, \bibinfo
  {author} {\bibfnamefont {H.}~\bibnamefont {Song}}, \bibinfo {author}
  {\bibfnamefont {Z.}~\bibnamefont {Zhao}}, \bibinfo {author} {\bibfnamefont
  {A.}~\bibnamefont {Minoofar}}, \bibinfo {author} {\bibfnamefont
  {R.}~\bibnamefont {Zhang}}, \bibinfo {author} {\bibfnamefont
  {H.}~\bibnamefont {Song}}, \bibinfo {author} {\bibfnamefont {H.}~\bibnamefont
  {Zhou}}, \bibinfo {author} {\bibfnamefont {X.}~\bibnamefont {Su}}, \bibinfo
  {author} {\bibfnamefont {C.}~\bibnamefont {Liu}}, \bibinfo {author}
  {\bibfnamefont {N.}~\bibnamefont {Hu}}, \bibinfo {author} {\bibfnamefont
  {M.}~\bibnamefont {Tur}},\ and\ \bibinfo {author} {\bibfnamefont {A.~E.}\
  \bibnamefont {Willner}},\ }\bibfield  {title} {\bibinfo {title} {Simulation
  of near-diffraction- and near-dispersion-free {OAM} pulses with controllable
  group velocity by combining multiple frequencies, each carrying a {B}essel
  mode},\ }\href@noop {} {\bibfield  {journal} {\bibinfo  {journal} {Opt.
  Lett.}\ }\textbf {\bibinfo {volume} {46}},\ \bibinfo {pages} {4678} (\bibinfo
  {year} {2021})}\BibitemShut {NoStop}%
\bibitem [{\citenamefont {Yessenov}\ \emph
  {et~al.}(2021{\natexlab{b}})\citenamefont {Yessenov}, \citenamefont {Free},
  \citenamefont {Chen}, \citenamefont {Johnson}, \citenamefont {Lavery},
  \citenamefont {Alonso},\ and\ \citenamefont {Abouraddy}}]{Yessenov21arx}%
  \BibitemOpen
  \bibfield  {author} {\bibinfo {author} {\bibfnamefont {M.}~\bibnamefont
  {Yessenov}}, \bibinfo {author} {\bibfnamefont {J.}~\bibnamefont {Free}},
  \bibinfo {author} {\bibfnamefont {Z.}~\bibnamefont {Chen}}, \bibinfo {author}
  {\bibfnamefont {E.~G.}\ \bibnamefont {Johnson}}, \bibinfo {author}
  {\bibfnamefont {M.~P.}\ \bibnamefont {Lavery}}, \bibinfo {author}
  {\bibfnamefont {M.~A.}\ \bibnamefont {Alonso}},\ and\ \bibinfo {author}
  {\bibfnamefont {A.~F.}\ \bibnamefont {Abouraddy}},\ }\bibfield  {title}
  {\bibinfo {title} {Space-time wave packets localized in all dimensions},\
  }\href@noop {} {\bibfield  {journal} {\bibinfo  {journal} {arXiv preprint
  arXiv:2111.03095}\ } (\bibinfo {year} {2021}{\natexlab{b}})}\BibitemShut
  {NoStop}%
\bibitem [{\citenamefont {Carpenter}\ \emph {et~al.}(2015)\citenamefont
  {Carpenter}, \citenamefont {Eggleton},\ and\ \citenamefont
  {Schr{\"o}der}}]{Carpenter15NP}%
  \BibitemOpen
  \bibfield  {author} {\bibinfo {author} {\bibfnamefont {J.}~\bibnamefont
  {Carpenter}}, \bibinfo {author} {\bibfnamefont {B.~J.}\ \bibnamefont
  {Eggleton}},\ and\ \bibinfo {author} {\bibfnamefont {J.}~\bibnamefont
  {Schr{\"o}der}},\ }\bibfield  {title} {\bibinfo {title} {Observation of
  {E}isenbud-{W}igner-{S}mith states as principal modes in multimode fibre},\
  }\href@noop {} {\bibfield  {journal} {\bibinfo  {journal} {Nat. Photon.}\
  }\textbf {\bibinfo {volume} {9}},\ \bibinfo {pages} {751} (\bibinfo {year}
  {2015})}\BibitemShut {NoStop}%
\bibitem [{\citenamefont {Carpenter}\ \emph {et~al.}(2017)\citenamefont
  {Carpenter}, \citenamefont {Eggleton},\ and\ \citenamefont
  {Schr{\"o}der}}]{Carpenter17LPR}%
  \BibitemOpen
  \bibfield  {author} {\bibinfo {author} {\bibfnamefont {J.}~\bibnamefont
  {Carpenter}}, \bibinfo {author} {\bibfnamefont {B.~J.}\ \bibnamefont
  {Eggleton}},\ and\ \bibinfo {author} {\bibfnamefont {J.}~\bibnamefont
  {Schr{\"o}der}},\ }\bibfield  {title} {\bibinfo {title} {Comparison of
  principal modes and spatial eigenmodes in multimode optical fibre},\
  }\href@noop {} {\bibfield  {journal} {\bibinfo  {journal} {Laser Photon.
  Rev.}\ }\textbf {\bibinfo {volume} {11}},\ \bibinfo {pages} {1600259}
  (\bibinfo {year} {2017})}\BibitemShut {NoStop}%
\bibitem [{\citenamefont {Caloz}\ and\ \citenamefont
  {Deck-L{\'e}ger}(2020)}]{Caloz20IEEE}%
  \BibitemOpen
  \bibfield  {author} {\bibinfo {author} {\bibfnamefont {C.}~\bibnamefont
  {Caloz}}\ and\ \bibinfo {author} {\bibfnamefont {Z.-L.}\ \bibnamefont
  {Deck-L{\'e}ger}},\ }\bibfield  {title} {\bibinfo {title} {Spacetime
  metamaterials--{P}art {I: G}eneral concepts},\ }\href@noop {} {\bibfield
  {journal} {\bibinfo  {journal} {IEEE Trans. Antennas Propag.}\ }\textbf
  {\bibinfo {volume} {68}},\ \bibinfo {pages} {1569} (\bibinfo {year}
  {2020})}\BibitemShut {NoStop}%
\bibitem [{\citenamefont {B{\'e}langer}(1986)}]{Belanger86JOSAA}%
  \BibitemOpen
  \bibfield  {author} {\bibinfo {author} {\bibfnamefont {P.~A.}\ \bibnamefont
  {B{\'e}langer}},\ }\bibfield  {title} {\bibinfo {title} {Lorentz
  transformation of packetlike solutions of the homogeneous-wave equation},\
  }\href@noop {} {\bibfield  {journal} {\bibinfo  {journal} {J. Opt. Soc. Am.
  A}\ }\textbf {\bibinfo {volume} {3}},\ \bibinfo {pages} {541} (\bibinfo
  {year} {1986})}\BibitemShut {NoStop}%
\bibitem [{\citenamefont {Longhi}(2004{\natexlab{b}})}]{Longhi04OE}%
  \BibitemOpen
  \bibfield  {author} {\bibinfo {author} {\bibfnamefont {S.}~\bibnamefont
  {Longhi}},\ }\bibfield  {title} {\bibinfo {title} {Gaussian pulsed beams with
  arbitrary speed},\ }\href@noop {} {\bibfield  {journal} {\bibinfo  {journal}
  {Opt. Express}\ }\textbf {\bibinfo {volume} {12}},\ \bibinfo {pages} {935}
  (\bibinfo {year} {2004}{\natexlab{b}})}\BibitemShut {NoStop}%
\bibitem [{\citenamefont {Saari}\ and\ \citenamefont
  {Reivelt}(2004)}]{Saari04PRE}%
  \BibitemOpen
  \bibfield  {author} {\bibinfo {author} {\bibfnamefont {P.}~\bibnamefont
  {Saari}}\ and\ \bibinfo {author} {\bibfnamefont {K.}~\bibnamefont
  {Reivelt}},\ }\bibfield  {title} {\bibinfo {title} {Generation and
  classification of localized waves by {L}orentz transformations in {F}ourier
  space},\ }\href@noop {} {\bibfield  {journal} {\bibinfo  {journal} {Phys.
  Rev. E}\ }\textbf {\bibinfo {volume} {69}},\ \bibinfo {pages} {036612}
  (\bibinfo {year} {2004})}\BibitemShut {NoStop}%
\bibitem [{\citenamefont {Qu}\ \emph {et~al.}(2016)\citenamefont {Qu},
  \citenamefont {Deck-L{\'e}ger}, \citenamefont {Caloz},\ and\ \citenamefont
  {Skorobogatiy}}]{Qu16JOSAA}%
  \BibitemOpen
  \bibfield  {author} {\bibinfo {author} {\bibfnamefont {H.}~\bibnamefont
  {Qu}}, \bibinfo {author} {\bibfnamefont {Z.-L.}\ \bibnamefont
  {Deck-L{\'e}ger}}, \bibinfo {author} {\bibfnamefont {C.}~\bibnamefont
  {Caloz}},\ and\ \bibinfo {author} {\bibfnamefont {M.}~\bibnamefont
  {Skorobogatiy}},\ }\bibfield  {title} {\bibinfo {title} {Frequency generation
  in moving photonic crystals},\ }\href@noop {} {\bibfield  {journal} {\bibinfo
   {journal} {J. Opt. Soc. Am. A}\ }\textbf {\bibinfo {volume} {33}},\ \bibinfo
  {pages} {1616} (\bibinfo {year} {2016})}\BibitemShut {NoStop}%
\end{thebibliography}%

%\bibliographyfullrefs{diffraction}

\end{document}